\documentclass[preprintnumbers,amssymb, nofootinbib, nobibnotes,aps,prd]{revtex4}
\usepackage{soul, color}
\usepackage{amsmath}
\usepackage{braket}
\usepackage{graphicx}% Include figure files
\usepackage{dcolumn}% Align table columns on decimal point
\usepackage{bm}% bold math

\newcommand{\beq}{\begin{equation}}
\newcommand{\eeq}{\end{equation}}
\newcommand{\bea}{\begin{eqnarray}}
\newcommand{\eea}{\end{eqnarray}}

\newcommand{\del}{\partial}

\newcommand{\half}{\frac{1}{2}}

\newcommand{\pa}{p_\alpha}
\newcommand{\pb}{p_\beta}

\newcommand{\pab}{\left(p_\alpha \cdot p_\beta\right)}

\begin{document}

\title{Gravitational radiation from the classical spinning double copy}
\author{Jingping Li}
\author{Siddharth G. Prabhu}
\affiliation{Physics Department, Yale University, New Haven, CT 06520, USA}
\date{\today}

\begin{abstract}
	We establish a correspondence between perturbative classical gluon and gravitational radiation emitted by spinning sources, to linear order in spin. This is an extension of the non-spinning classical perturbative double copy and uses the same color-to-kinematic replacements. The gravitational  theory has a scalar (dilaton) and a 2-form field (the Kalb-Ramon axion) in addition to the graviton. In arXiv:1712.09250, we computed axion radiation in the gravitational theory to show that the correspondence fixes its action. Here, we present complete details of the gravitational computation. In particular, we also calculate the graviton and dilaton amplitudes in this theory and find that they precisely match with the predictions of the double copy. This constitutes a non-trivial check of the classical double copy correspondence, and brings us closer to the goal of simplifying the calculation of gravitational wave observables for astrophysically relevant sources.
\end{abstract}
\maketitle

\section{Introduction}
\label{sec:intro}
Einstein's theory of General Relativity, one of the most beautiful triumphs of modern physics, describes classical gravity to the best of our knowledge. However, the computational effort required to solve Einstein's equations, even perturbatively, is significant. %The recent discovery of gravitational waves by LIGO \cite{TheLIGOScientific:2017qsa} underscores a pressing need for efficient ways of calculating high precision observables. 
On the other hand, in recent years, we have seen a series of remarkable developments in the study of perturbative scattering amplitudes in quantum field theory with both theoretical and practical significance. One could ask whether these methods are useful for the problem of obtaining solutions of classical gravity.

A recent promising approach in this direction relies on an idea first discovered in the context of quantum scattering amplitudes in gauge and gravity theories by Bern, Carrasco and Johansson (BCJ)~\cite{Bern:2008qj,Bern:2010ue,Bern:2010yg}.
%BCJ discovered that when written in a certain form, perturbative gauge theory amplitudes can be transformed into their corresponding gravitational counterparts via a simple color-kinematic substitution. 
%They showed that diagrammatic numerators for perturbative gravity amplitudes can be obtained as products of the corresponding numerators of gauge theories.
Numerators of gauge theory Feynman diagrams factorize into color factors (arising from the gauge group) and kinematic factors (made up of velocities, polarizations, etc). Simply put, the BCJ prescription is to write the gauge theory amplitude in a certain form and replace every color factor with its kinematic factor counterpart. This procedure then gives the corresponding gravity amplitude. This BCJ \emph{double copy} was, in turn, motivated by the closed string-open string amplitude relations due to Kawai, Lewellen and Tye (KLT)~\cite{Kawai:1985xq}. KLT showed that the integrands of closed tree-level string amplitudes factorize into open string ones. In the field theory limit, these express gravity tree amplitudes as a product of two corresponding gauge theory tree amplitudes. The BCJ double copy includes the field theory limit of the KLT relations as a special case. It has been proven for all tree-level scattering amplitudes \cite{Bern:2010yg} and there is increasing evidence at the loop level in various settings \cite{Loops}. See \cite{Carrasco:2015iwa} for a recent review.

The question of whether the double copy extends to classical solutions in gauge theory and General Relativity was first raised in \cite{Monteiro:2014cda}. Their method of obtaining solutions of General Relativity in the Kerr-Schild gauge was extended and studied in more detail in \cite{KS2, KS3, KSothers}. The BCJ double copy allows for the calculation of precision observables in gravity that were previously not amenable to a direct computation, by replacing such a calculation with the analogous gauge theory computation. Can such an idea be used to simplify the perturbative expansion of the equations of General Relativity and reduce the computational effort required for gravitational wave calculations?\footnote{It is to be noted that an effective field theory approach to tackle the binary inspiral problem was introduced in \cite{Goldberger:2004jt} and extended to include spinning sources in \cite{Porto:2005ac}. A recent comprehensive review can be found in \cite{Porto:2016pyg}.} 

Goldberger and Ridgway probed this question \cite{Goldberger:2016iau} by starting with a system of well-separated point color charges coupled to the Yang-Mills field. They calculated the Yang-Mills radiation that the sources generate, by self-consistently solving the equations of motion for the sources and the field perturbatively. Remarkably, they found that a set of simple color-to-kinematic replacements produces gravitational radiation emitted by an analogous system of point masses. It was shown in \cite{Goldberger:2017frp} that these color-to-kinematic replacement rules can also be used to generate Yang-Mills radiation from scalar radiation, thereby completing a two-fold classical double copy for leading order radiation. This classical perturbative double copy was extended to radiation from sources in time-dependent orbits in \cite{Goldberger:2017vcg}, such as the bound orbits relevant for gravitational wave detection \cite{TheLIGOScientific:2017qsa}. Ref. \cite{Chester:2017} showed that Einstein-Yang Mills radiation can be obtained from Yang-Mills scalar radiation. Another approach to generate space times perturbatively, that is inspired by the double copy, can be found in  \cite{Luna:2016hge}.

In this paper, we complete the extension of the perturbative classical double copy to the case of radiation from spinning sources started in \cite{Goldberger:2017axi}. Our goal is to compute the gravitational radiation emitted from a system of spinning sources moving on general time-dependent trajectories, that satisfy the equations of motion, in $d$ dimensions.  The motion of extended objects under a gravitational field has been approached through a variety of ways \cite{Mathisson:1937zz,Papapetrou:1951pa,Dixon:1970zza,Hanson:1974qy,Bailey:1975fe}. The formalism we use to describe spinning objects is detailed in the appendices of \cite{Goldberger:2017axi}, and is equivalent to the one used in \cite{Porto:2005ac,Porto+Rothstein}, in the context of effective field theories for extended gravitational sources. 

Instead of attempting to solve the Einstein's equations with spinning sources, we look to utilize the classical double copy~\cite{Goldberger:2016iau}. To this end, we consider, instead, a system of point colored charges, with color variable $c_a(\tau) $ \cite{sikivie}, that couple to the Yang-Mills field.\footnote{Finite size corrections are systematically accounted for by including higher order terms in an effective field theory framework, see \cite{Goldberger:2004jt}.} Each point charge possesses a spin angular momentum $S^{\mu\nu}(\tau)$ which couples to the Yang-Mills field via a chromomagnetic spin dipole coupling 
\begin{equation}
S_{int}={g_s\kappa\over 2} \int d\tau \, c_a \, S^{\mu\nu} F^a_{\mu\nu},
\end{equation}
with coupling strength $\kappa$, and $\tau$ the worldline coordinate. 
We let the particles evolve self-consistently under their equations of motion and compute, to linear order in spin, the amplitude of Yang-Mills radiation ${\cal A}^{\mu}_a(k) $ that they generate. We then employ the simple color-to-kinematics substitutions \cite{Goldberger:2016iau,Goldberger:2017vcg} 
to get a double copy radiation amplitude ${\cal A}^{\mu\nu}(k)$,
\beq
\epsilon^{*a}_\mu(k)  {\cal A}^{\mu}_a(k) \mapsto \epsilon^{*}_\mu(k) {\tilde\epsilon}^{*}_\nu(k) {{\cal A}}^{\mu\nu}(k).
\eeq
Consistency of the double copy amplitude ${{\cal A}}^{\mu\nu}(k)$ with gravitational Ward identities sets the chromomagnetic dipole coupling strength $\kappa$ for each particle to be the same constant $\kappa=-1$. The double copy amplitude ${\cal A}^{\mu\nu}(k)$ can, in general, be decomposed into its antisymmetric, symmetric-traceless and trace components. The corresponding radiation fields are also expected by decomposing products of vector irreducible representations of the massless little group $SO(d-2)$
%\begin{align}
%A_\mu\otimes A_\nu = \phi\oplus h_{\mu\nu}\oplus B_{\mu\nu},
%\end{align}
\begin{align}
n\otimes n = &1\oplus\frac{n(n+1)}{2}-1\oplus\frac{n(n-1)}{2} \\
&\phi \hspace{10mm}h_{\mu\nu}  \hspace{18mm} B_{\mu\nu}
\end{align}
where $\phi$ is a scalar (dilaton), $ h_{\mu\nu}$ the graviton, and $B_{\mu\nu}$ the Kalb-Ramond axion \cite{Kalb:1974yc}.\footnote{To be explicit, we refer to the 2-form field $B_{\mu\nu}$ in any dimension as the \emph{axion}.} 

In the case of non-spinning sources, the double copied field is symmetric, thereby implying the field content of the gravitational theory to be $(h_{\mu \nu}, \phi)$ . This can be understood by noting that one cannot write down a linear interaction of non-spinning particles with the axion field. Alternately, in this case, gravitational radiation can be seen as arising as a two-fold double copy of the bi-adjoint scalar radiation \cite{Goldberger:2017frp}. The latter theory enjoys a $G\times\tilde G$ global symmetry and is invariant under exchange of these two groups. The color-kinematic substitution rules, that take an adjoint index of each group to a Lorentz index, treat both adjoint indices corresponding to these two groups symmetrically. Hence, the resulting gravitational radiation is symmetric under exchange of the Lorentz indices. The action of the gravitational theory was shown to be \cite{Goldberger:2016iau}
\beq
\label{eq:nonspinaction}
S= -2 m_{Pl}^{d-2}\int d^d x \sqrt{g} \left[R -(d-2) g^{\mu\nu}\partial_\mu\phi\partial_\nu\phi\right]-\sum_\alpha m_\alpha \int d\tau e^\phi.
\eeq

For spinning sources, we expect the field content of the gravitational theory to be $(h_{\mu \nu},B_{\mu \nu}, \phi)$. Decomposing the double copy amplitude lands us at graviton, dilaton, and axion radiation in this theory. We write down the most general action with two derivatives using diffeomorphism invariance and 2-form gauge invariance. Consistency with the double copy fixes the action to be 
\begin{equation}
\label{eq:sb}
S_g= -2 m_{Pl}^{d-2}\int d^d x \sqrt{g} \left[R -(d-2) g^{\mu\nu}\partial_\mu\phi\partial_\nu\phi + {1\over 12} e^{-4\phi} H_{\mu\nu\sigma} H^{\mu\nu\sigma} \right],
\end{equation}
where $H_{\mu\nu\sigma} = (d B)_{\mu\nu\sigma}$ is the field strength of the 2-form.
This action also describes the BCJ double copy of pure gluons  \cite{Bern:2010yg} (see also~\cite{Luna:2016hge}) and appears in the low energy effective action of oriented closed strings. Compared to the non-spinning case, the spinning sources have an additional interaction, namely that with the axion field given by
\begin{equation}
\label{eq:SBaction}
S_{HS}=\frac{1}{4}  \int dx^\mu   H_{\mu\nu\sigma} S^{\nu\sigma}e^{-2\phi}.
\end{equation}
We note that this action differs from the one in \cite{Goldberger:2017axi} as the "string frame" metric ${\tilde g}_{\mu\nu}=g_{\mu\nu}e^{2\phi}$ was used to define spin there, as opposed to the ordinary metric $g_{\mu\nu}$ used in this paper (for more details, refer to Sec. \ref{sec:Grav}). 

The rest of the paper is organised as follows. In section \ref{sec:YM}, we review the computation of classical gluon radiation from a system of several spinning sources to leading order in spin that was obtained in \cite{Goldberger:2017axi}. We obtain the double copy of this radiation amplitude in section \ref{sec:DC} and decompose it into radiation in graviton, dilaton and axion channels. In section \ref{sec:Grav}, we calculate the corresponding radiation amplitudes emitted by a collection of several spinning masses in the gravitational theory given by Eqs. (\ref{eq:sb},\ref{eq:SBaction}). We discuss our results and further questions raised in section \ref{sec:Disc}.

\section{Yang-Mills radiation}
\label{sec:YM}
We begin by reviewing the calculation of Yang-Mills radiation emitted by a classical system of several spinning colored particles in $d$ dimensions, presented in \cite{Goldberger:2017axi}. 
For each particle, with worldline coordinate s, the degrees of freedom are a worldline trajectory $x^{\mu}(s)$, a spin angular momentum $S^{\mu \nu}(s)$, and a color charge $c^{a}(s)$ \cite{sikivie} transforming in the adjoint representation of the gauge group $G$. We first present some details of the spinning formalism of \cite{Goldberger:2017axi}, that are needed to describe the interactions of such a system with a gauge field.\footnote{We use the conventions $D_\mu = \partial_\mu + i g_s A^a_\mu T^a$,  $[T^a,T^b]=if^{abc} T^c$, $(T_{\mbox{\tiny{adj}}}^a)^b_c=-if_{abc}$.}

In order to describe the spin degree of freedom, we endow each worldline with an orthonormal reference frame $e^I_{\mu}(s)$ \cite{Hanson:1974qy}.
The spin $S^{IJ}(s)$ is then introduced as the variable conjugate to the angular velocity 
\begin{equation}
\Omega^{IJ} \equiv \eta^{\mu\nu} e^I_\mu {d\over ds}    e^J_\nu = -\Omega^{JI},
\end{equation}
whereas the momentum $p_I(s)$ is conjugate to $x_I(s)$. 
In $d$ dimensions, the number of spatial rotational degrees of freedom is $\frac{1}{2} (d-1)(d-2)$. Hence, we need to impose a constraint on the spin angular momentum to get this physical number of degrees of freedom. Following \cite{Hanson:1974qy,Pryce:1948sp}, this can be done in a number of different ways. We choose to impose the constraint that the spin is transverse to the momentum,
\beq
S^{\mu \nu}p_\nu=0. \label{eq:constraint}
\eeq 
We also introduce an einbein $e$ that enforces worldline reparametrization invariance ($e(s) ds$ is invariant under $s \mapsto s'(s) )$, and a Lagrange multiplier $\lambda_I $ that enforces the spin constraint above.

Each particle is described by the action
\begin{align}
\label{eq:Spp}
S_{pp} =&-\int dx^\mu e^I_\mu p_I + {1\over 2}\int ds e\left(p_I p^I - m^2(S) + \cdots\right)+{1\over 2} \int ds S^{IJ} \Omega_{IJ}  + \int ds e \lambda_I S^{IJ} p_J \nonumber\\
& - g_s\int dx^\mu  c_a A^a_\mu +{g_s\kappa\over 2} \int ds e  c_a S^{\mu\nu} F^a_{\mu\nu},
\end{align}
where the first line has all the terms that describe a free particle, and the second line contains the interaction terms of the particle with the gauge field. Here, $g_s$ is the Yang-Mills coupling constant, and $\kappa$ is the spin dipole coupling constant.
This action, together with the usual Yang-Mills action in the bulk, constitutes the complete action for the system of particles interacting with a gauge field. 

The resulting equations of motion are the following. Varying the action with respect to the gauge field, we have the usual Yang-Mills field equations
\begin{equation}
\label{eq:YM}
D_\nu F^{\nu\mu}_a(x) = g_s J_a^\mu(x),
\end{equation}
with the color current generated by the particles given by
\begin{equation}
J_{a}^{\mu}(x)\equiv-{1\over g_s} {\delta\over \delta A^a_\mu(x)} S_{pp} =\sum_{\alpha}\int dx^{\mu}_{\alpha}c_{\alpha}^{a}\frac{\delta^{d}(x-x_{\alpha}(s_{\alpha}))}{\sqrt{g}}-\kappa_\alpha\int ds_{\alpha}e_{\alpha}S^{\mu\nu}D_{\nu}\bigg[c_{\alpha}^{a}\frac{\delta^{d}(x-x_{\alpha}(s_{\alpha}))}{\sqrt{g}}\bigg].
\end{equation}
Here, the sum runs over all the particles, indexed by $\alpha$.

Imposing current conservation covariantly, $D_{\mu}J_a^{\mu}=0$, gives the equation of motion in color space,
\begin{equation}
\left(\dot{x}_\alpha\cdot D\right)c_{\alpha}^{a}=\frac{i\kappa_{\alpha}g_s}{2}e_{\alpha}S_{\alpha}^{\mu\nu}\left[F_{\mu\nu},c_{\alpha}\right]^{a}.\label{eq:col}
\end{equation}
The energy momentum tensor for a single spinning particle comes out to be 
\begin{equation}
T_{pp}^{\mu\nu}(x) \equiv-{2\over \sqrt{g}} {\delta\over \delta g_{\mu\nu}(x)}S_{pp}=\int dx^{(\mu} p^{\nu)} {\delta(x-x(s))} + \int  dx^{(\mu}S^{\nu)\sigma}  \partial_\sigma {\delta(x-x(s))} -{\kappa g_s} \int ds {\delta(x-x(s))\over \sqrt{g}} c_a {F^a}_\sigma{}^{(\mu}{} S^{\nu)\sigma},
\end{equation}
where the brackets $()$ indicate symmetrization of the corresponding indices. The integral of the divergence of conserved currents with arbitrary support $X$ should vanish on-shell $\int d^d x\sqrt{g}X_\nu\nabla_{\mu} \left(T^{\mu\nu}_{YM} + T^{\mu\nu}_{pp}\right)\bigr\vert_{\text{on-shell}}=0$. This leads to the equations of motion for the momentum and the spin,

\begin{equation} 
\frac{d}{ds}p_{\alpha}^{\mu}=g_s c_{\alpha}^{a}F_{a}^{\mu\nu}\dot{x}_{\alpha\nu}-\frac{\kappa_{\alpha}g_s e_{\alpha}}{2}S_{\alpha}^{\rho\sigma}c_{\alpha}^{a}D^{\mu}F_{\rho\sigma}^{a},\label{eq:mom}
\end{equation}
\begin{equation}
\frac{d}{ds}S_{\alpha}^{\mu\nu}=\dot{x}_{\alpha}^{\nu}p_{\alpha}^{\mu}-\dot{x}_{\alpha}^{\mu}p_{\alpha}^{\nu}-\kappa_{\alpha} g_s e_{\alpha}c_{\alpha}^{a}F_{\sigma}^{a\ \mu}S_{\alpha}^{\nu\sigma}+\kappa_{\alpha}g_s e_{\alpha}c_{\alpha}^{a}F_{\sigma}^{a\ \nu}S_{\alpha}^{\mu\sigma}.\label{eq:sp}
\end{equation}
The motion of the particles is thus described by the system of equations Eqs.~(\ref{eq:col},\ref{eq:mom},\ref{eq:sp}), subject to the constraint Eq.~(\ref{eq:constraint}). Alternately, these equations of motion can also be obtained by varying the action with respect to $(x^\mu,e^I_\mu,e,p_I,S^{IJ},\lambda_I)$. The constants of the motion are $c_a c^a$, $S_{\mu \nu} S^{\mu \nu}$ and $m^2 = p_\mu p^\mu + g_s\kappa c_a S^{\mu\nu}F^a_{\mu\nu}$. 

From the invariance of the spin constraint 
\begin{equation}
\frac{d}{ds}\left(S^{\mu\nu}p_\nu\right)=0,\label{eq:pos}
\end{equation}
we can solve for the velocity $v^\mu\equiv \dot{x}^\mu$ in terms of the other variables. In the following, we use reparametrization freedom to choose $e_\alpha$ such that $s_\alpha=\tau_\alpha$, the proper time for each particle, whereby $p^\mu_\alpha\simeq v_\alpha^\mu$ up to $O(S^0)$.

In the Lorenz gauge, $\partial_\mu A^\mu_a=0$, the Yang-Mills field equations Eq. (\ref{eq:YM}) take the form
\beq
\label{eq:gcurr}
\Box A^\mu_a\equiv g_s{\tilde J}^\mu_a(x)=g_s J^\mu_a + g_s f^{abc} A^b_\nu(\partial^\nu A_c^\mu - F_c^{\mu\nu}),
\eeq
defining the source current ${\tilde J}^\mu_a(x)$, which includes contributions from both the point sources as well as the field configuration. It is conserved, $\partial_\mu {\tilde J}^\mu_a(x)=0$,
and related to observables measured at null infinity. The specific relation between the radiation field at null infinity, and the source current in momentum space ${\tilde J}^\mu_a(k)=\int d^d x e^{ik\cdot x} {\tilde J}^\mu_a(x)$ depends on the dimension $d$. For example, in $d=4$ dimensions, the radiation field is given by
\beq
\label{eq:lda}
\lim_{r\rightarrow\infty}  A_a^\mu(x)  = {g_s\over 4\pi r}\int {d\omega\over 2\pi} e^{-i\omega t} {\tilde J}^\mu_a(k),
\eeq
with $k^\mu = (\omega,{\vec k})=\omega(1,{\vec x}/r)$. In any dimensions, the total energy-momentum radiated out to infinity in polarization channel $r$ is given by
\begin{equation}
\Delta P_\text{r}^\mu=\int_k (2\pi)\theta (k^0)\delta (k^2)k^\mu |\epsilon^{*a}_{\text{r},\nu}(k)g_s\tilde{J}_a^\nu(k)|^2,\label{eq:pw}
\end{equation}
 with $\epsilon^{a}_{\text{r},\mu}(k)$ being gluon polarization vectors. These are  normalized as $\epsilon^{*a}_{\text{r}}(k)\cdot\epsilon^{b}_{\text{r}'}(k)=-\delta_{ab}\delta_{\text{r}\text{r}'}$, and satisfy the gauge condition $k\cdot\epsilon_{\text{r}}^a(k)=0$. (The polarization indices do not play any role in our calculations, so they will be suppressed from now on). Suitable integrals of the momentum space source current ${\tilde J}^\mu_a(k)$ thus produce physical observables at null infinity. Hence, in what follows, our object of interest is ${\tilde J}^\mu_a(k)$. We compute it perturbatively %\textcolor{red}{write about power counting} 
in the Yang-Mills coupling constant,\footnote{See \cite{Goldberger:2016iau,Goldberger:2017frp,Goldberger:2017vcg} for a detailed discussion of the explicit small expansion parameter that suppresses higher order contributions.} and to linear order in spin, by consistently solving the system of equations for the particles and the field.

%
%
%\begin{equation}
%c_{\alpha}^{a}(s\rightarrow-\infty)=c_{\alpha}^{a},\label{eq:coli}
%\end{equation}
%\begin{equation}
%p_{\alpha}^{\mu}(s\rightarrow-\infty)=p_{\alpha}^{\mu},\label{eq:momi}
%\end{equation}
%\begin{equation}
%S_{\alpha}^{\mu\nu}(s\rightarrow-\infty)=S_{\alpha}^{\mu},\label{eq:spi}
%\end{equation}
%\begin{equation}
%x_{\alpha}^{\mu}(s\rightarrow-\infty)=b_{\alpha}^{\mu}+p_{\alpha}^{\mu}s,\label{eq:posi}
%\end{equation}
%which are the solutions to the non-interacting equations of motion.
%At finite parameter time, we encode the deflections of higher orders
%from the interactions as
%\begin{equation}
%c_{\alpha}^{a}(s)=c_{\alpha}^{a}+\bar{c}_{\alpha}^{a}(s),\label{eq:col}
%\end{equation}
%\begin{equation}
%p_{\alpha}^{\mu}(s)=p_{\alpha}^{\mu}+\bar{p}_{\alpha}^{\mu}(s),\label{eq:mom}
%\end{equation}
%\begin{equation}
%S_{\alpha}^{\mu\nu}(s)=S_{\alpha}^{\mu\nu}+\bar{S}_{\alpha}^{\mu\nu}(s),\label{eq:sp}
%\end{equation}
%\begin{equation}
%x_{\alpha}^{\mu}(s)=b_{\alpha}^{\mu}+p_{\alpha}^{\mu}s+z_{\alpha}^{\mu}(s).\label{eq:pos}
%\end{equation}
%To remove the redundant Lorentz indices, we employ the same set of shorthand notations as in []
%\begin{align*}
%S_{\alpha}^{\mu\nu}p_{\nu} & \equiv (S_\alpha \wedge p)^{\mu},
%\end{align*}
%\[
%k_{\mu}S_{\alpha}^{\mu\nu}p_{\nu}\equiv (k\wedge p)_\alpha.
%\]
%for any vectors $p$ and $q$. In these notations,
%the source at lowest order linear in spin is
\begin{figure}
	\centering
	\includegraphics[scale=0.59]{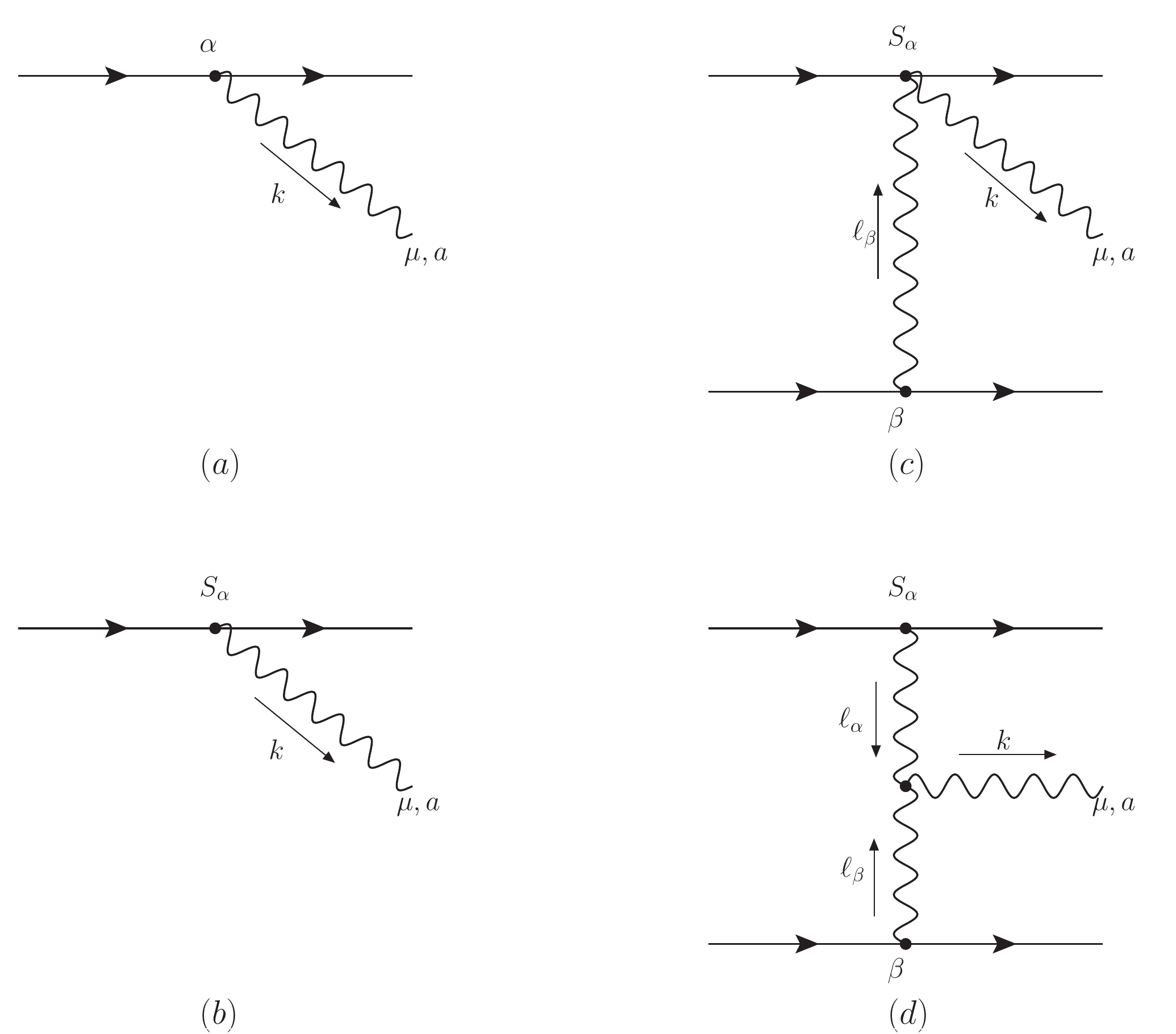}
	\caption{\label{fig:gluon1pt} Feynman diagrams that contribute to leading order gluon radiation.  Diagram $(a)$ corresponds to the spin-independent contribution to the source current $\tilde{J}^\mu_a(k)$. Diagrams $(b)$-$(d)$ correspond to spin-dependent contributions to the source current.}    
\end{figure}

In the following, we employ the same notation as in \cite{Goldberger:2017axi},  
\begin{align}
S_{\alpha}^{\mu\nu}p_{\nu} & \equiv(S_{\alpha}\wedge p)^{\mu},
\end{align}
\begin{equation}
k_{\mu}S_{\alpha}^{\mu\nu}p_{\nu}\equiv(k\wedge p)_{\alpha}.
\end{equation}
to denote contractions of the spin angular momentum with any Lorentz vectors $k$ and $p$. We also use $\mathcal{O}(\ldots)$ notation to denote contributions at a particular order. 

The leading order current can be seen as the contributions of the  Feynman diagrams in Figs.~\ref{fig:gluon1pt}(a) and (b) to lowest order in the coupling constant. Following the results in \cite{Goldberger:2017vcg}, we can work with particles travelling along general time-dependent (possibly bound) orbits, say $x^\mu_\alpha(\tau_{\alpha})$, $p^\mu_\alpha(\tau_{\alpha})$, $ c^a_\alpha(\tau_{\alpha})$, $S^{\mu\nu}_\alpha(\tau_{\alpha})$. In the following, we drop the explicit dependence on $\tau$, so that, for example, $x^\mu_\alpha\equiv x^\mu_\alpha(\tau_{\alpha})$. 
Then to all orders in perturbation, the contribution from the sum of these diagrams can be written as
\begin{equation}
\tilde{J}^\mu_a(k)\bigr\vert_{\text{Fig.~\ref{fig:gluon1pt}(a)+(b)}}= \sum_\alpha\int d\tau_\alpha\, e^{ik\cdot x_\alpha}c^a_\alpha \left[{v}^\mu_\alpha  +i \kappa_\alpha (S_\alpha\wedge k)^\mu\right].\label{eq:pc}
\end{equation}
At leading order, this gives rise to the field\footnote{As we are in a classical setup, our propagators, here and in the rest of the paper, are assumed to be defined with retarded boundary conditions $1/k^2 = 1/((k^0+i\epsilon)^2 - {\vec k}^2)$ and $1/k\cdot p=1/(k\cdot p +i\epsilon)$.}
\begin{equation}
\label{eq:lofield}
A_a^\mu(x)\bigr\vert_{\mathcal{O}(g_s^1)} = g_s \int {d^d\ell\over (2\pi)^d} {e^{-i\ell\cdot x}\over\ell^2} \tilde{J}^\mu_a(\ell)\bigr\vert_{\mathcal{O}(g_s^1)} = g_s \sum_\alpha\int d\tau_\alpha \, {d^d\ell\over (2\pi)^d} {e^{-i\ell\cdot (x- x_\alpha)}\over\ell^2} c^a_\alpha \left[{p}^\mu_\alpha +i \kappa_\alpha (S_\alpha\wedge\ell)^\mu \right].
\end{equation}
This lowest order field then induces corrections in color, position, momentum, and spin of the particles which causes the particle to radiate at the next order in perturbation.
The first contribution to radiation comes from gluons emitted directly by the particles. This is the $\mathcal{O}(g_s^2)$ contribution to Eq. (\ref{eq:pc}), given by
\begin{subequations}
\begin{align}
\tilde{J}^\mu_a(k)&\bigr\vert_{\text{Fig.~\ref{fig:gluon1pt}(a)+(b)};\mathcal{O}(g_{s}^{2},S^1)}= \sum_\alpha\int d\tau_\alpha e^{ik\cdot x_\alpha}{i\over k\cdot v_\alpha} \left[ {\dot c}^a_\alpha {v}^\mu_\alpha+c^a_\alpha \left\{{\dot v}^\mu_\alpha - {k\cdot {\dot v}_\alpha\over k\cdot v_\alpha} {v}^\mu_\alpha \right\}\right]\Biggr\vert_{\mathcal{O}(g_{s}^{2},S^1)}\label{line1}\\
& - \sum_\alpha \kappa_\alpha \int d\tau_\alpha e^{ik\cdot x_\alpha}{1\over k\cdot v_\alpha} \left[ {\dot c}^a_\alpha  (S_\alpha\wedge k)+c^a_\alpha\left\{ ({\dot S}_\alpha\wedge k)- {k\cdot {\dot v}_\alpha\over k\cdot v_\alpha} (S_\alpha\wedge k)^\mu \right\}\right]\Biggr\vert_{\mathcal{O}(g_{s}^{2},S^1)}, \label{line2}
\end{align}
\end{subequations}
%\begin{align}
%\tilde{J}_{a}^{\mu}(k)\vert_{\text{Fig.};\mathcal{O}(g^2,S)}=&\sum_{\alpha}e^{ik\cdot b_{\alpha}}[p_{\alpha}^{\mu}\bar{c}_{\alpha}^{a}(\omega)\vert_{\mathcal{O}(g^2,S)}-i\omega c_{\alpha}^{a}z_{\alpha}^{\mu}(\omega)\vert_{\mathcal{O}(g^2,S)}+ip_{\alpha}^{\mu}c_{\alpha}^{a}k\cdot z_{\alpha}(\omega)\vert_{\mathcal{O}(g^2,S)}\label{line1}\\
%&+i\kappa_{\alpha}k_{\nu}c_{\alpha}^{a}\bar{S}_{\alpha}^{\mu\nu}(\omega)\vert_{\mathcal{O}(g^2,S)}+i\kappa_{\alpha}k_{\nu}S_{\alpha}^{\mu\nu}\bar{c}_{\alpha}^{a}(\omega)\vert_{\mathcal{O}(g^2,S^0)}-\kappa_{\alpha}k_{\nu}S_{\alpha}^{\mu\nu}c_{\alpha}^{a}k\cdot z_{0\alpha}(\omega)\vert_{\mathcal{O}(g^2,S^0)}]_{\omega=k\cdot p_{\alpha}},\label{line2}
%\end{align}
where the first line corresponds to inserting the spin-dependent solutions \cite{Goldberger:2017axi} while the second line corresponds to inserting spin-independent solutions \cite{Goldberger:2016iau}. These substitutions give 
\begin{align}
\text{(\ref{line1})}  &=ig_s^{2}\sum_{{\alpha,\beta\atop \alpha\neq\beta}}\int d\mu_{\alpha\beta}(k)\bigg[[c_{\alpha},c_{\beta}]^{a}\frac{\ell_{\alpha}^{2}}{k\cdot p_{\alpha}}\bigg\{\kappa_{\alpha}(\ell_{\beta}\wedge p_{\beta})_{\alpha}p_{\alpha}^{\mu}-\kappa_{\text{\ensuremath{\beta}}}(\ell_{\beta}\wedge p_{\alpha})_{\beta}p_{\alpha}^{\mu}\bigg\}\nonumber \\
& +(c_{\alpha}\cdot c_{\beta})c_{\alpha}^{a}\bigg\{(1+\kappa_{\alpha})\frac{\ell_{\alpha}^{2}}{p_{\alpha}^{2}}\bigg[(p_{\alpha}\cdot p_{\beta})(S_{\alpha}\wedge \ell_{\beta})^{\mu}-(k\cdot p_{\alpha})(S_{\alpha}\wedge p_{\beta})^{\mu}+(k\wedge p_{\beta})_{\alpha}p_{\alpha}^{\mu}-\frac{p_{\alpha}\cdot p_{\beta}}{k\cdot p_{\alpha}}(\ell_{\alpha}\wedge \ell_{\beta})_{\alpha}p_{\alpha}^{\mu}\bigg]\nonumber \\
& +\kappa_{\beta}\frac{\ell_{\alpha}^{2}}{k\cdot p_{\alpha}}\bigg[(k\cdot p_{\alpha})(S_{\beta}\wedge \ell_{\beta})^{\mu}+(\ell_{\beta}\wedge \ell_{\alpha})_{\beta}p_{\alpha}^{\mu}+(\ell_{\beta}\wedge p_{\alpha})_{\beta}\bigg(\ell_{\beta}^{\mu}-\frac{k\cdot \ell_{\beta}}{k\cdot p_{\alpha}}p_{\alpha}^{\mu}\bigg)\bigg]\nonumber\\
& -\kappa_{\alpha}\frac{\ell_{\alpha}^{2}}{k\cdot p_{\alpha}}\bigg[(\ell_{\beta}\wedge p_{\beta})_{\alpha}\bigg(\ell_{\beta}^{\mu}-\frac{k\cdot \ell_{\beta}}{k\cdot p_{\alpha}}p_{\alpha}^{\mu}\bigg)\bigg]\bigg\}\bigg],\label{eq:A} 
\end{align}
\begin{align}
\text{(\ref{line2})} &=ig_s^{2}\sum_{{\alpha,\beta\atop \alpha\neq\beta}}\int d\mu_{\alpha\beta}(k)\bigg[[c_{\alpha},c_{\beta}]^{a}\frac{p_{\alpha}\cdot p_{\beta}}{k\cdot p_{\alpha}}\ell_{\alpha}^{2}(S_{\alpha}\wedge k)^{\mu}\nonumber \\
& +(c_{\alpha}\cdot c_{\beta})c_{\alpha}^{a}\bigg\{(1+\kappa_{\alpha})\frac{\ell_{\alpha}^{2}}{p_{\alpha}^{2}}\bigg[(p_{\alpha}\cdot p_{\beta})(S_{\alpha}\wedge \ell_{\beta})^{\mu}-(k\cdot p_{\alpha})(S_{\alpha}\wedge p_{\beta})^{\mu}+(k\wedge p_{\beta})_{\alpha}p_{\alpha}^{\mu}-\frac{p_{\alpha}\cdot p_{\beta}}{k\cdot p_{\alpha}}(\ell_{\alpha}\wedge \ell_{\beta})_{\alpha}p_{\alpha}^{\mu}\bigg]\nonumber \\
& +\kappa_{\alpha}\frac{\ell_{\alpha}^{2}}{k\cdot p_{\alpha}}\Big[(\ell_{\alpha}\wedge \ell_{\beta})_{\alpha}p_{\beta}^{\mu}-(k\wedge p_{\beta})_{\alpha}\ell_{\beta}^{\mu}-(k\cdot p_{\beta})(S_{\alpha}\wedge \ell_{\beta})^{\mu}+(k\cdot \ell_{\beta})(S_{\alpha}\wedge p_{\beta})^{\mu}\Big]\nonumber\\
& +\frac{\ell_{\alpha}^{2}}{k\cdot p_{\alpha}}\bigg[\frac{p_{\alpha}\cdot p_{\beta}}{k\cdot p_{\alpha}}(k\cdot \ell_{\beta})-(k\cdot p_{\beta})\bigg](S_{\alpha}\wedge k)^{\mu}\bigg\}\bigg], \label{eq:B}
\end{align}
where we have introduced an integration measure over worldline parameters and momenta given by
\beq
\label{eq:mudef}
d\mu_{\alpha\beta}(k) \equiv d\tau_\alpha d\tau_\beta \left[{d^d\ell_\alpha\over (2\pi)^d} {e^{i\ell_\alpha\cdot x_\alpha}\over\ell^2_\alpha}\right] \left[{d^d\ell_\beta\over (2\pi)^d}{e^{i\ell_\beta\cdot x_\beta}\over\ell^2_\beta}\right] (2\pi)^d \delta^d(k-\ell_\alpha-\ell_\beta).
\eeq

At this order $O(g_s^2)$ in perturbation, there are two contributions from diagrams without deflections in the particle trajectories. The first of these is from Fig.~\ref{fig:gluon1pt}(c),
\begin{equation}
\tilde{J}_{a}^{\mu}(k)\bigr\vert_{\text{Fig.~\ref{fig:gluon1pt}(c)};\mathcal{O}(g_{s}^{2},S^1)}=ig_s^{2}\sum_{{\alpha,\beta\atop \alpha\neq\beta}}\int d\mu_{\alpha\beta}(k)[c_{\alpha},c_{\beta}]^{a}\ell_{\alpha}^{2}(S_{\alpha}\wedge p_{\beta})^{\mu}.\label{eq:C}
\end{equation}
The second of the zero deflection contributions is from the diagram with the triple vertex in Fig.~\ref{fig:gluon1pt}(d),
\begin{align}
\tilde{J}_{a}^{\mu}(k)\bigr\vert_{\text{Fig.~\ref{fig:gluon1pt}(d)};\mathcal{O}(g_{s}^{2},S^1)} & =-ig_s^{2}\sum_{{\alpha,\beta\atop \alpha\neq\beta}}\int d\mu_{\alpha\beta}(k)[c_{\alpha},c_{\beta}]^{a}[2(k\cdot p_{\beta})(S_{\alpha}\wedge \ell_{\alpha})^{\mu}+(\ell_{\alpha}\wedge p_{\beta})_{\alpha}(\ell_{\alpha}-\ell_{\beta})^{\mu}+2(\ell_{\alpha}\wedge \ell_{\beta})_{\alpha}p_{\beta}^{\mu}].\label{eq:D}
\end{align}

We can write down the total expression for the leading order radiation (as written in \cite{Goldberger:2017axi}) coming from spinning particles in general orbits consistent with the equations of motion. The result is a sum of two color structures
\begin{eqnarray}
{\tilde J}^\mu_a(k)\bigr\vert_{{\cal O}(g_s^2,S^1)} =i g_s^2  \sum_{\alpha,\beta\atop \alpha\neq\beta}\int d\mu_{\alpha\beta}(k) \left[(c_\alpha\cdot c_\beta) c^a_\alpha {\cal A}^\mu_s+  [c_\alpha,c_\beta]^a {\cal A}_{adj}^\mu\right],
\end{eqnarray}
with
\begin{eqnarray}
\label{eq:aadj}
\nonumber
{\cal A}^\mu_{adj} &\equiv& \kappa_\alpha \left[(\ell_\alpha\wedge p_\beta)_\alpha (\ell_\beta-\ell_\alpha)^\mu -  {\ell_\alpha^2\over k\cdot p_\alpha}  (\ell_\beta\wedge p_\beta)_\alpha p_\alpha^\mu -{\ell_\beta^2\over k\cdot p_\beta}  (\ell_\alpha\wedge p_\beta)_\alpha p_\beta^\mu + \ell_\alpha^2 (S_\alpha\wedge p_\beta)^\mu\right]\\
\nonumber\\
& & {}  -2 \kappa_\alpha(k\cdot p_\beta) \left[(S_\alpha\wedge \ell_\alpha)^\mu - {(k\wedge\ell_\alpha)_\alpha\over k\cdot p_\beta} p^\mu_\beta\right]
- \kappa_\alpha {\ell_\alpha^2\over k\cdot p_\alpha} (p_\alpha\cdot p_\beta) (S_\alpha\wedge k)^\mu.
\end{eqnarray}
and
\begin{eqnarray}
\label{eq:as}
\nonumber
{\cal A}_s^\mu &\equiv&    {(1+\kappa_\alpha)^2 \over m_\alpha^2}\ell_\alpha^2  \left[  (k\cdot p_\alpha)\left\{(S_\alpha\wedge p_\beta)^\mu - {(k\wedge p_\beta)_\alpha\over k\cdot p_\alpha} p_\alpha^\mu\right\} + (p_\alpha\cdot p_\beta)\left\{(S_\alpha\wedge \ell_\beta)^\mu - {(k\wedge\ell_\beta)_\alpha\over k\cdot p_\alpha} p_\alpha^\mu\right\} \right]\\
\nonumber\\
\nonumber
& & -\kappa_\beta\ell_\alpha^2\left[(S_\beta\wedge\ell_\beta)^\mu- {(k\wedge\ell_\beta)_\beta\over k\cdot p_\alpha} p_\alpha^\mu \right]\\
\nonumber\\
\nonumber 
& & {} + \kappa_\alpha^2  { \ell_\alpha^2 \over k\cdot p_\alpha} \left[(k\cdot p_\beta)\left\{(S_\alpha\wedge \ell_\beta)^\mu - {(k\wedge\ell_\beta)_\alpha\over k\cdot p_\beta}  p_\beta^\mu \right\}  - (k\cdot\ell_\beta)\left\{ (S_\alpha\wedge p_\beta)^\mu - {(k\wedge p_\beta)_\alpha\over k\cdot p_\beta} \ell_\beta^\mu\right\}\right]\\
\nonumber\\
\nonumber
& &  {} + \kappa_\alpha {\ell_\alpha^2\over k\cdot p_\alpha}\left[(\ell_\beta\wedge p_\beta)_\alpha\left\{\ell_\beta^\mu - {k\cdot\ell_\beta\over k\cdot p_\alpha} p_\alpha^\mu\right\}   +(k\cdot p_\beta) (S_\alpha\wedge k)^\mu \right] +\kappa_\beta  {\ell_\alpha^2\over k\cdot p_\alpha} (\ell_\beta\wedge p_\alpha)_\beta\left[\ell_\beta^\mu - {k\cdot\ell_\beta\over k\cdot p_\alpha} p^\mu_\alpha\right]\\
& & {} -   \kappa_\alpha {\ell_\alpha^2\over (k\cdot p_\alpha)^2} (p_\alpha\cdot p_\beta) (k\cdot\ell_\beta) (S_\alpha\wedge k)^\mu
\end{eqnarray}
It is easy to check that this result satisfies the Ward identity $k_\mu {\tilde J}_a^\mu(k)=0$ even off-shell.

\section{Double Copy}
\label{sec:DC}
As in the spinless case \cite{Goldberger:2016iau,Goldberger:2017vcg}, we transform the Yang-Mills radiation by the following set of color-kinematic substitution rules
%\begin{eqnarray}
%\label{eq:dcrules}
%c^a_\alpha &\mapsto&   p_\alpha^\mu,\nonumber\\
%\left[ c_\alpha, c_\beta\right]^a  &\mapsto& \Gamma^{\mu\nu\rho}(-k,\ell_\alpha,\ell_\beta) p_{\alpha\nu} p_{\beta\rho}= {1\over 2}\left[(p_\alpha\cdot p_\beta) (\ell_\beta-\ell_\alpha)^\mu + p_\beta\cdot (\ell_\alpha+k) p_\alpha^\mu-  p_\alpha\cdot (\ell_\beta+k) p_\beta^\mu \right],
%\end{eqnarray}
\begin{eqnarray}
\label{eq:dcrules}
{c}^a_\alpha(\tau) &\mapsto& p_\alpha^\mu(\tau), \\  
 \left[{c}_\alpha(\tau), {c}_\beta(\tau)\right]^a & \mapsto & \Gamma^{\mu\nu\rho}(-k,\ell_\alpha,\ell_\beta) p_{\nu\alpha}(\tau) p_{\rho\beta}(\tau),
\end{eqnarray}
where $\Gamma^{\mu\nu\rho}(-k,\ell_\alpha,\ell_\beta)$ is the kinematic part of the 3-point gluon vertex Feynman rule, defined as  $$\Gamma^{\mu\nu\rho}(-k,\ell_\alpha,\ell_\beta) \equiv {1\over 2}\left[ (\ell_\beta-\ell_\alpha)^\mu \eta^{\nu \rho} + (\ell_\alpha+k)^{\rho} \eta^{\mu \nu} -  (\ell_\beta+k)^{\nu} \eta^{\mu \rho}\right].$$ We also identify the respective coupling constants 
\bea \label{eq:ccdc}
g_s\mapsto \frac{1}{2m^{(d-2)/2}_{Pl}}\label{coupling}\equiv\eta,
\eea
 In the non-spinning case, the momenta remained unchanged under the double copy $p^\mu_\alpha(\tau) \mapsto p^\mu_\alpha(\tau)$. Similarly, in the spinning case, we use the substitution $S^{\mu\nu}_\alpha(\tau) \mapsto S^{\mu\nu}_\alpha(\tau)$.  We use the above substitution rules to transform the Yang-Mills radiation amplitude ${\cal A}^\mu_a(k)$, defined as ${\cal A}^\mu_a(k)\equiv g_s {\tilde J}^{\mu}_a(k)\bigr\vert_{k^2=0}$, and obtain the double copy radiation amplitude ${{\cal A}}^{\mu\nu}(k)$, with $k^2=0$, as 
\beq
\epsilon^{*a}_\mu(k)  {\cal A}^{\mu}_a(k) \mapsto \epsilon^{*}_\mu(k) {\tilde\epsilon}^{*}_\nu(k) {{\cal A}}^{\mu\nu}(k),\label{eq:dcp}
\eeq
where the on-shell gluon polarization $\epsilon^{*a}_\mu(k)$ has been formally replaced by a product of on-shell independent polarizations $\epsilon^{*}_\mu(k) {\tilde\epsilon}^{*}_\nu(k) $. Thus, the double copy amplitude ${\cal A}^{\mu\nu}(k)$ is defined up to terms that vanish when dotted into these polarization vectors. Explicitly, it is given by 
\begin{eqnarray}
\label{eq:Amunu}
\nonumber
{\cal A}^{\mu\nu}(k)\bigr\vert_{\mathcal{O}(\eta^3, S^1)} &=& {i\over 8 m_{Pl}^{(d-2)/2}} \sum_{\alpha,\beta\atop \alpha\neq\beta} \int d\mu_{\alpha\beta}(k)\left[\left( {1\over 2} (p_\alpha\cdot p_\beta) (\ell_\beta-\ell_\alpha)^\nu + (k\cdot p_\beta) p_\alpha^\nu-  (k\cdot p_\alpha) p_\beta^\nu\right) {\cal A}^\mu_{adj}\right. \\
& &\hspace{5cm} {}\left. - (p_\alpha\cdot p_\beta) {p^\nu_\alpha} {{\cal A}}^\mu_{s}\right],
\end{eqnarray}
where ${\cal A}^\mu_{adj}$ and ${\cal A}^\mu_s$ are given in Eqs.~(\ref{eq:aadj}),~(\ref{eq:as}) respectively. 

We see that $k_\mu  {\tilde J}^{\mu}(k)=0$ automatically guarantees  $k_\mu {{\cal A}}^{\mu\nu}(k)=0$ because the color-kinematic substitution rules do not affect this Lorentz index. For ${\cal A}^{\mu\nu}(k)$ to define the radiation amplitudes consistently in a gravitational theory, we also need it to satisfy the Ward identity in the second Lorentz index $k_\nu {{\cal A}}^{\mu\nu}(k)=0$. Unlike the non-spinning case, this now imposes an extra constraint on the Yang-Mills theory \cite{Goldberger:2017axi}, namely that  
\beq \kappa_{\alpha}=-1. \eeq 
Thus, while the Yang-Mills theory is consistent for any value of the chromomagnetic coupling constant $\kappa_{\alpha}$, we find that the double copy procedure only works when all the particles couple to the gauge field with this special value of the coupling constant. As was noted in \cite{Goldberger:2017axi}, in $d=4$, this value corresponds to the so-called \emph{natural} value \cite{Ferrara:1992yc,Holstein:2006ry} of the gyromagnetic ratio $g=2$. 
%So, though our particles are classical, which means the spins take continuous values and have magnitudes $|{\vec S}|\gg \hbar$
For this special value of the coupling constant, we can write the double copy amplitude as

\bea
\label{eq:dc}
&&{\cal A}^{\mu\nu}(k)\bigr\vert_{\mathcal{O}(\eta^3, S^1)}=-\frac{i}{8m_{Pl}^{3(d-2)/ 2}}\sum_{\alpha,\beta\atop \alpha\neq\beta}\int d\mu_{\alpha\beta}(k) \left[  \frac{\ell_{\alpha }^2 \left(  p_{\alpha } \cdot p_{\beta }\right)p_{\alpha }^{\nu }  }{k \cdot  p_{\alpha } } \left\{\frac{k \cdot  \ell_{\beta }}{k \cdot  p_{\alpha }} \left( \left( p_{\alpha } \cdot p_{\beta }\right)  \left(S_{\alpha }\wedge k \right)^{\mu }+p_{\alpha }^{\mu } \left[\left(\ell_{\beta }\wedge p_{\beta }\right)_{\alpha } \right.\right.\right.\right. \nonumber \\
& &  \left.\left.\left. -\left(\ell_{\beta }\wedge p_{\alpha }\right)_{\beta }\right] \right) -\left(k \cdot  \ell_{\beta }\right) \left(S_{\alpha }\wedge p_{\beta }\right)^{\mu } + \frac{1}{2} \left(\ell_{\beta }^{\mu }-\ell_{\alpha }^{\mu }\right) \left[ \left( \ell_{\alpha } \wedge p_{\beta }\right)_{\alpha } + \left( \ell_{\beta }\wedge  p_{\alpha }\right) _{\beta }\right]- p_{\alpha }^{\mu }\left(\ell_{\alpha }\wedge \ell_{\beta }\right) _{\beta } -p_{\beta }^{\mu}\left(\ell_{\alpha }\wedge \ell_{\beta }\right) _{\alpha }  \right.\nonumber \\
& &-  \left(k \cdot  p_{\beta }\right) \left(S_{\alpha }\wedge \ell_{\alpha } \right)^{\mu }\left. +\left(k \cdot  p_{\alpha }\right) \left(S_{\beta }\wedge \ell_{\beta } \right)^{\mu }\right\}+\left\{ p_{\alpha }^{\nu } \left(k \cdot  p_{\beta }\right)- p_{\beta }^{\nu } \left(k \cdot  p_{\alpha }\right)+\half \left(\ell_{\beta }^{\nu }-\ell_{\alpha }^{\nu }\right) \left(  p_{\alpha } \cdot p_{\beta } \right)\right\}
  \nonumber \\
& &\left\{-\frac{\ell_{\alpha }^2}{k \cdot  p_{\alpha }} \left(\left(  p_{\alpha } \cdot p_{\beta } \right) \left(S_{\alpha }\wedge k \right)^{\mu }\left.-p_{\alpha }^{\mu } \left[\left( \ell_{\beta }\wedge p_{\alpha }\right)_{\beta } -\left( \ell_{\beta }\wedge p_{\beta }\right) _{\alpha }\right]\right)+\ell_{\beta }^{\mu } \left[ \left(\ell_{\beta }\wedge p_{\alpha }\right) _{\beta }+\left( \ell_{\alpha }\wedge p_{\beta }\right)_{\alpha }\right]\right.\right. \nonumber \\
& &-2 p_{\beta }^{\mu } (\ell_{\alpha }\wedge \ell_{\beta })_{\alpha } \left. \left.+ 2 \left(k \cdot  p_{\alpha }\right) \left(S_{\beta }\wedge \ell_{\beta } \right)^{\mu }+\ell_{\alpha }^2 \left(S_{\alpha }\wedge p_{\beta } \right)^{\mu } \right\} \right],
\eea
where we have used the gauge freedom to add a term proportional to $k^\mu$, so that, on shell, the double copy satisfies the Ward identity in both indices. 
%We have also left out terms with more total derivatives; for instance, terms proportional to $\ell_\alpha\cdot p_\alpha$ is equivalent to an additional total derivative on the Fourier kernel.
We can use the double copy to obtain radiation amplitudes in various channels, by decomposing the product of polarizations in Eq.~(\ref{eq:dcp}) into irreducible representations of the massless little group $SO(d-2)$ as
\begin{equation}
\epsilon_\mu\tilde{\epsilon}_\nu=\epsilon_{\mu\nu}+a_{\mu\nu}+\frac{\epsilon\cdot\tilde{\epsilon}}{d-2}\pi_{\mu\nu},
\end{equation}
where $\epsilon_{\mu\nu}\equiv\frac{1}{2}(\epsilon_{\mu}\tilde{ \epsilon}_{\nu}+\epsilon_{\nu}\tilde{ \epsilon}_{\mu})-\frac{\epsilon\cdot\tilde{\epsilon}}{d-2}\pi_{\mu\nu}$, $a_{\mu\nu}\equiv\frac{1}{2}(\epsilon_{\mu}\tilde{ \epsilon}_{\nu}-\epsilon_{\nu}\tilde{ \epsilon}_{\mu})$ and $\frac{\epsilon\cdot\tilde{\epsilon}}{d-2}\pi_{\mu\nu}$ are the symmetric-traceless, the antisymmetric and the trace parts respectively. Here $\pi_{\mu\nu}\equiv\eta_{\mu \nu}-\frac{(k_\mu q_\nu+k_\nu q_\mu)}{k.q}$ is the projector onto the $(d-2)$ dimensional space spanned by the polarization vectors orthogonal to both the external momentum $k$, and an arbitrary reference vector $q$, satisfying $q^2=0$. 

In the non-spinning case, it was shown, by an explicit computation, that the double copy produces non-vanishing radiation amplitudes in the dilaton and graviton channels of a gravitational theory whose action is $S=S_g+S_{pp}$, with
\beq
\label{eq:dilgravbulk}
S_g= -2 m_{Pl}^{d-2}\int d^d x \sqrt{g} \left[R -(d-2) g^{\mu\nu}\partial_\mu\phi\partial_\nu\phi\right],
\eeq 
and 
\beq
\label{eq:dilgravpp}
S_{pp}= -\sum_\alpha m_\alpha \int d\tau_{\alpha} e^\phi.
\eeq 
In $d=4$, for example, this means that the radiation fields at null infinity calculated in this dilaton gravity theory, can be reproduced with the double copy, by writing
\begin{eqnarray}
h_{\pm}(t,{\vec n}) &=&  {4G_N\over r} \int {d\omega\over 2\pi} e^{-i\omega t} \epsilon^*{}_{\pm}^{\mu\nu}(k) {\cal A}_{\mu\nu}(k),\\
\phi(t,{\vec n}) &=&  {G_N\over r} \int {d\omega\over 2\pi} e^{-i\omega t} \eta^{\mu\nu} {\cal A}_{\mu\nu}(k),
\end{eqnarray}
Here ${\vec n}={\vec k}/|{\vec k}|$, and $\omega=k^0$ is the frequency of radiation (in $d=4$,  $G_N=1/32\pi m_{Pl}^2$). 

As discussed in the introduction, we expect the double copy amplitude to also have a non-vanishing antisymmetric component. This corresponds to the Kalb-Ramond axion $B_{\mu \nu}(x)$ in the gravitational theory. Hence, our purported gravitational theory has the field content $(h_{\mu \nu},B_{\mu \nu}, \phi)$. We now write down the radiation amplitudes in this theory, as predicted by the double copy. 

The axion amplitude is defined to be
\beq 
{\hat{\cal A}}_{B}(k)\equiv a^*_{\mu\nu}(k) {\cal A}^{\mu\nu}(k).
\eeq
2-form gauge invariance implies that the polarization tensor is defined up to gauge transformations $a_{\mu\nu}(k)\rightarrow a_{\mu\nu}(k) + k_\mu \zeta_\nu(k) - k_\nu \zeta_\mu(k)$ for an arbitrary gauge parameter $\zeta_\nu(k)$.
The double copy predicts the axion radiation amplitude to be
\bea
& &{\hat{\cal A}}_{B}(k)\bigr\vert_{\mathcal{O}(\eta^3, S^1)}=-\frac{i a^*_{\mu\nu}(k) }{16m_{Pl}^{3(d-2)/ 2}}\sum_{\alpha,\beta\atop \alpha\neq\beta}\int d\mu_{\alpha\beta}(k) \left[ \left(\frac{\ell_{\alpha }^2 \left(  p_{\alpha } \cdot p_{\beta } \right) }{k \cdot  p_{\alpha } }\left( p_{\alpha }^{\nu }  \left\{k \cdot  \ell_{\beta } \left(\frac{ p_{\alpha } \cdot p_{\beta }  }{k \cdot  p_{\alpha }}\left(S_{\alpha }\wedge k \right)^{\mu }- \left( S_{\alpha }\wedge p_{\beta }\right)^{\mu }\right)\right. \right.\right.\right. \nonumber \\
& & \left.\left. -\left(k \cdot  p_{\beta }\right) \left(S_{\alpha }\wedge k\right)^{\mu } -  \left(k \cdot  p_{\beta }\right) \left(S_{\alpha }\wedge \ell_{\alpha } \right)^{\mu } + \frac{1}{2} \left(\ell_{\beta }^{\mu }-\ell_{\alpha }^{\mu }\right) \left(k\wedge p_{\beta }\right)_{\alpha }- p_{\beta }^{\mu } \left(\ell_{\alpha }\wedge \ell_{\beta }\right)_{\alpha }\right\}\right) \nonumber \\
& &\left.-  \half \left(  p_{\alpha } \cdot p_{\beta } \right) \left(\ell_\beta-\ell_\alpha\right)^{\nu}\left(S_{\alpha }\wedge k\right)^{\mu } \right) +\ell_{\alpha }^2 \left( \pab \left(\pa^\nu \left(S_{\beta }\wedge \ell_{\beta } \right)^{\mu }+ \pb^\nu \left(S_{\alpha }\wedge k \right)^{\mu }  -p_{\beta }^{\mu }\left(\ell_\beta-\ell_\alpha\right)^{\nu}\left(\ell_{\alpha }\wedge \ell_{\beta }\right)_{\alpha }  \right) \right.\nonumber\\
& &\left.- p_{\beta }^{\nu } p_{\alpha }^{\mu } \left[\left(\ell_{\beta }\wedge p_{\alpha }\right)_{\beta } -\left( \ell_{\beta }\wedge p_{\beta }\right)_{\alpha }\right]\right) +\left( p_{\alpha }^{\nu } \left(k \cdot  p_{\beta }\right)- p_{\beta }^{\nu } \left(k \cdot  p_{\alpha }\right)+\half \left(\ell_{\beta }^{\nu }-\ell_{\alpha }^{\nu }\right) \left(  p_{\alpha } \cdot p_{\beta } \right)\right)
\left( 2 \left(k \cdot  p_{\alpha }\right) \left(S_{\beta }\wedge \ell_{\beta } \right)^{\mu }\right. \nonumber \\
& &\left.\left. +\ell_{\alpha }^2 \left(S_{\alpha }\wedge p_{\beta } \right) ^{\mu }\right) +\ell_{\beta }^{\mu }\left(\left(k \cdot  p_{\beta }\right)p_{\alpha }^{\nu } -\left(k \cdot  p_{\alpha }\right)p_{\beta }^{\nu }\right) \left[\left( \ell_{\alpha }\wedge p_{\beta }\right)_{\alpha } + \left( \ell_{\beta }\wedge p_{\alpha }\right)_{\beta }\right]-\{\mu \leftrightarrow \nu\} \right]. \label{eq:bfielddc}
\eea

We define the dilaton amplitude to be
\beq {\hat{\cal A}}_{s}(k)\equiv \frac{1}{(d-2)^{1/2}}\eta_{\mu \nu} \, \mathcal{A}^{\mu \nu} (k),
\eeq
the double copy prediction for which is
\begin{align}
\hat{\mathcal{A}}_{s}(k)\bigr\vert_{\mathcal{O}(\eta^{3},S^{1})}&=-\frac{i}{8m_{Pl}^{3(d-2)/2}(d-2)^{1/2}}\sum_{{\alpha,\beta\atop \alpha\neq\beta}}\int d\mu_{\alpha\beta}(k)p_{\alpha}^{2}\bigg[\frac{(p_{\alpha}\cdot p_{\beta})(k\cdot\ell_{\beta})\ell_{\alpha}^{2}}{(k\cdot p_{\alpha})^{2}}\Big\{(\ell_{\beta}\wedge p_{\alpha})_{\beta}-(\ell_{\beta}\wedge p_{\beta})_{\alpha}\Big\}\nonumber\\
&+\frac{\ell_{\alpha}^{2}}{k\cdot p_{\alpha}}\Big\{(p_{\alpha}\cdot p_{\beta})(k\wedge\ell_{\beta})_{\beta}+(k\cdot p_{\beta})(\ell_{\beta}\wedge p_{\beta})_{\alpha}-(k\cdot p_{\beta})(\ell_{\beta}\wedge p_{\alpha})_{\beta}\Big\}+2(k\cdot p_{\beta})(\ell_{\alpha}\wedge\ell_{\beta})_{\beta}\bigg].\label{eq:dilatondc}
\end{align}

Finally, the graviton amplitude is given by
\begin{equation}
\hat{\mathcal{A}}_{g}(k)\equiv \epsilon^*_{\mu\nu}\mathcal{A}^{\mu\nu}(k),\label{eq:gamp}
\end{equation}
where the polarization tensor $\epsilon_{\mu\nu}(k)$ is defined up to gauge transformations $\epsilon_{\mu\nu}(k)\rightarrow \epsilon_{\mu\nu}(k) + k_\mu \zeta_\nu(k) + k_\nu \zeta_\mu(k)$ for an arbitrary gauge parameter $\zeta_\nu(k)$. This predicts the total graviton amplitude to be
\begin{align}
\hat{\mathcal{A}}_{g}(k)\bigr\vert_{\mathcal{O}(\eta^{3},S^{1})} & =\frac{i\epsilon_{\mu\nu}^{*}}{8m_{Pl}^{3(d-2)/2}}\sum_{{\alpha,\beta\atop \alpha\neq\beta}}\int d\mu_{\alpha\beta}(k)\bigg[\frac{(p_{\alpha}\cdot p_{\beta})(k\cdot\ell_{\beta})\ell_{\alpha}^{2}}{2(k\cdot p_{\alpha})^{2}}\Big\{\Big((\ell_{\beta}\wedge p_{\alpha})_{\beta}-(\ell_{\beta}\wedge p_{\beta})_{\alpha}\Big)p_{\alpha}^{\mu}p_{\alpha}^{\nu}\nonumber \\
& -(p_{\alpha}\cdot p_{\beta})(S_{\alpha}\wedge k)^{\mu}p_{\alpha}^{\nu}\Big\}+\frac{p_{\alpha}\cdot p_{\beta}}{2(k\cdot p_{\alpha})}\bigg\{\frac{1}{2}\ell_{\alpha}^{2}(p_{\alpha}\cdot p_{\beta})(S_{\alpha}\wedge k)^{\mu}(\ell_{\beta}-\ell_{\alpha})^{\nu}+\ell_{\alpha}^{2}(k\cdot p_{\beta})(S_{\alpha}\wedge\ell_{\alpha})^{\mu}p_{\alpha}^{\nu}\nonumber \\
& +\ell_{\alpha}^{2}(k\cdot p_{\beta})(S_{\alpha}\wedge k)^{\mu}p_{\alpha}^{\nu}+\ell_{\alpha}^{2}(k\cdot\ell_{\beta})(S_{\alpha}\wedge p_{\beta})^{\mu}p_{\alpha}^{\nu}+\ell_{\alpha}^{2}(\ell_{\alpha}\wedge\ell_{\beta})_{\alpha}p_{\alpha}^{\mu}p_{\beta}^{\nu}+\ell_{\alpha}^{2}(\ell_{\alpha}\wedge\ell_{\beta})_{\beta}p_{\alpha}^{\mu}p_{\alpha}^{\nu}\nonumber \\
& -(\ell_{\alpha}\cdot\ell_{\beta})(\ell_{\alpha}\wedge p_{\beta})_{\alpha}k^{\mu}p_{\alpha}^{\nu}-\ell_{\alpha}^{2}(k\wedge p_{\beta})_{\alpha}\ell_{\beta}^{\mu}p_{\alpha}^{\nu}+\ell_{\alpha}^{2}\Big\{(\ell_{\beta}\wedge p_{\beta})_{\alpha}-(\ell_{\beta}\wedge p_{\alpha})_{\beta}\Big\}\bigg(2\ell_{\beta}-\frac{1}{2}k\bigg)^{\mu}p_{\alpha}^{\nu}\bigg\}\nonumber \\
& +\frac{(k\cdot p_{\beta})\ell_{\alpha}^{2}}{2(k\cdot p_{\alpha})}\Big\{(\ell_{\beta}\wedge p_{\beta})_{\alpha}-(\ell_{\beta}\wedge p_{\alpha})_{\beta}\Big\} p_{\alpha}^{\mu}p_{\alpha}^{\nu}+\frac{1}{2}(p_{\alpha}\cdot p_{\beta})\bigg\{-\ell_{\alpha}^{2}(S_{\alpha}\wedge k)^{\mu}p_{\beta}^{\nu}-\ell_{\alpha}^{2}(S_{\beta}\wedge\ell_{\beta})^{\mu}p_{\alpha}^{\nu}\nonumber \\
& +\bigg((k\cdot p_{\beta})(S_{\alpha}\wedge\ell_{\alpha})^{\mu}-\frac{1}{2}\ell_{\alpha}^{2}(S_{\alpha}\wedge p_{\beta})^{\mu}+(\ell_{\alpha}\wedge\ell_{\beta})_{\alpha}p_{\beta}^{\mu}+\frac{1}{2}(\ell_{\alpha}\wedge p_{\beta})_{\alpha}(\ell_{\alpha}-\ell_{\beta})^{\mu}\bigg)(\ell_{\beta}-\ell_{\alpha})^{\nu}\bigg\}\nonumber \\
& -\frac{\ell_{\alpha}^{2}}{2}\Big\{(\ell_{\beta}\wedge p_{\beta})_{\alpha}-(\ell_{\beta}\wedge p_{\alpha})_{\beta}\Big\} p_{\alpha}^{\mu}p_{\beta}^{\nu}+\bigg\{(k\cdot p_{\beta})(S_{\alpha}\wedge\ell_{\alpha})^{\mu}-\frac{1}{2}(S_{\alpha}\wedge p_{\beta})^{\mu}\nonumber \\
& +(\ell_{\alpha}\wedge\ell_{\beta})_{\alpha}p_{\beta}^{\mu}+(\ell_{\alpha}\wedge p_{\beta})_{\alpha}\bigg(\ell_{\alpha}-\frac{1}{2}k\bigg)^{\mu}\bigg\}\Big((k\cdot p_{\beta})p_{\alpha}^{\nu}-(k\cdot p_{\alpha})p_{\beta}^{\nu}\Big)+(\mu\leftrightarrow\nu)\bigg].\label{eq:gravitondc}
\end{align}

\section{The gravitational theory}
\label{sec:Grav}
In this section, we calculate the amplitudes for dilaton, graviton and axion radiation emitted by a set of spinning sources coupled to gravity. We note that the color-kinematic substitution only contributes to additional powers of momenta in the numerators so it can only improve the analyticity of the amplitude. Hence, we expect the resulting gravitational theory to be local. The most general action in the bulk, up to two derivatives, with field content $(h_{\mu \nu},B_{\mu \nu}, \phi)$ that preserves the symmetries, namely diffeomorphism invariance and 2-form gauge invariance, is
\begin{equation}
S_g= -2 m_{Pl}^{d-2}\int d^d x \sqrt{g} \left[R -(d-2) g^{\mu\nu}\partial_\mu\phi\partial_\nu\phi + {1\over 12} f(\phi) H_{\mu\nu\sigma} H^{\mu\nu\sigma} \right],
\end{equation}
where $H_{\mu\nu\sigma} = (d B)_{\mu\nu\sigma}$ is the field strength corresponding to the 2-form field and $f(\phi)=1 + f'(0)\phi+\cdots $. 

Next, we move to the point particle action. We calculate Yang-Mills radiation to linear order in spin, whereas, on the gravitational side, it is easy to see that the leading order interaction of gravitons with the spin is second order in spin. However, the double copy predicts a non-vanishing axion amplitude linear in spin. This suggests that the corresponding gravity theory has a linear interaction of the spin with the 2-form field. In \cite{Goldberger:2017axi}, we wrote this unique leading order interaction term as
\begin{equation}
\label{eq:ppb}
S_{HS} = \int dx^\mu  {\tilde\kappa}(\phi) H_{\mu\nu\sigma} \tilde{S}^{\nu\sigma},
\end{equation}
for an arbitrary function ${\tilde \kappa}(\phi)={\tilde\kappa} + {\tilde\kappa}'\phi+\cdots$. We used $\tilde{S}^{\mu\nu}=S^{IJ}\tilde{e}_I^\mu\tilde{e}_J^\nu$, with $\tilde{e}_I^\mu$ being the vielbeins defined with respect to the string frame metric, i.e. $\eta_{IJ}\tilde{e}_\mu^I \tilde{e}_\nu^J={\tilde g}_{\mu\nu}=g_{\mu\nu}e^{2\phi}$. The double copied field does not have any free parameters. This means that consistency with the double copy should fix all the unknown parameters on the gravitational side. Indeed, comparing the axion radiation amplitude in this theory to the double copy prediction, these parameters were fixed to be~\cite{Goldberger:2017axi}
 \begin{eqnarray}
f'(0) = -4, & \tilde{\kappa}'=0, & {\tilde \kappa} ={1\over 4}.
\end{eqnarray}
The bulk action is then given by
\begin{equation}
S_{g}=-2m_{Pl}^{d-2}\int d^{d}x\sqrt{g}\bigg[R-(d-2)g^{\mu\nu}\partial_{\mu}\phi\partial_{\nu}\phi+\frac{e^{-4\phi}}{12}H_{\mu\nu\rho}H^{\mu\nu\rho}\bigg].\label{eq:grav}
\end{equation}
In this paper, we define the spin via ${S}^{\mu\nu}=S^{IJ}{e}_I^\mu{e}_J^\nu$, where the vielbeins $e_{\mu}^{I}$ are defined with respect to the ordinary metric,  $\eta_{IJ}{e}_\mu^I {e}_\nu^J={ g}_{\mu\nu}$. Then, the complete worldline action for a single particle is
\begin{equation}
S_{pp}=\int ds\bigg[-\dot{x}^{\mu}e_{\mu}^{I}p_{I}e^{\phi}+\frac{1}{2}S^{IJ}\Omega_{IJ}+\frac{1}{2}e(p^{I}p_{I}-m^{2})e^{\phi}+e\lambda_{I}S^{IJ}p_{J}+\frac{1}{4}S^{\mu\nu}\dot{x}^{\sigma}H_{\mu\nu\sigma}e^{-2\phi}\bigg].\label{eq:pp}
\end{equation}
Here, the angular velocity is defined with a covariant derivative, $\Omega_{IJ}=g_{\mu\nu}e^\mu_I \frac{D}{Ds}e^\nu_J\equiv e^\mu_I \dot{x}_\rho \nabla^\rho e^\nu_J$.
Note that though the unknown dilaton dependent functions in the action above have been written as exponentials, our computation only really fixes these functions to linear order in the dilaton. We expect the complete bulk action to be given by Eq.~(\ref{eq:grav}) as it describes the BCJ double copy of pure Yang-Mills \cite{Bern:2015uc}. It also arises as the leading low energy effective theory of the common sector of oriented string theories \cite{Scherk:1974mc,Gross:1987hs}.
 
 In the next subsection, we find the equations of motion for the system of fields and particles. We work with the ordinary metric and later, fix the worldline parameter $s$ to be the proper time per unit mass, $s=\tau$. This ensures that we get the non-spinning action and equations of motion in \cite{Goldberger:2016iau} when the spin is set to zero and particle masses are restored. To get the relevant results in \cite{Goldberger:2017axi}, we only need to switch back to the conformally rescaled metric $\tilde{g}_{\mu \nu}=g_{\mu\nu}e^{2\phi}$ and reparametrize the worldline coordinate to be the conformal proper time per unit mass, $s=\tilde{\tau} e^{-\phi}$. Of course, worldline reparametrization invariance ensures the invariance of the total amplitude. In subsection \ref{sec:LO}, we work out the leading order fields and the changes they induce in the momenta, color and spin of the particles. We use these field values and particle deflections to calculate the leading order axion, graviton, and dilaton radiation in subsections \ref{sec:axirad}, \ref{sec:gravrad}, and \ref{sec:dilrad} respectively.
 
\subsection{Equations of motion and solutions}
The equation of motion for the dilaton is 
\begin{equation} 
\nabla^\mu\nabla_\mu\phi-\frac{e^{-4\phi}}{6(d-2)}H_{\mu\nu\rho}H^{\mu\nu\rho}=-\frac{1}{4m_{Pl}^{d-2}(d-2)}J,\label{eq:eomd}
\end{equation}
where we have defined the source term on the RHS to be
\begin{equation}
J\equiv\sum_{\alpha}\int ds\bigg(\dot{x}_{\alpha}^{\mu}p_{\alpha\mu}e^{\phi}+S_{\alpha}^{\mu\nu}\dot{x}_{\alpha}^{\sigma}H_{\mu\nu\sigma}e^{-2\phi}\bigg)\frac{\delta^{d}(x-x_{\alpha})}{\sqrt{g}}.\label{eq:sd}
\end{equation}
We also derive the equation of motion of the axion to be
\begin{equation}
\nabla_{\lambda}(e^{-4\phi}H^{\mu\nu\lambda})=\frac{1}{m_{Pl}^{d-2}}J^{\mu\nu},\label{eq:eomh}
\end{equation}
whose source term is defined to be 
\begin{equation}
J^{\mu\nu}\equiv\sum_{\alpha}\frac{1}{4}\int ds(S_{\alpha}^{\lambda\mu}\dot{x}_{\alpha}^{\nu}+S_{\alpha}^{\nu\lambda}\dot{x}_{\alpha}^{\mu}+S_{\alpha}^{\mu\nu}\dot{x}_{\alpha}^{\lambda})\nabla_{\lambda}\bigg[e^{-2\phi}\frac{\delta^{d}(x-x_{\alpha})}{\sqrt{g}}\bigg].\label{eq:Bsource}
\end{equation}

We now find the equations of motion for the particles using the same method as in \cite{Goldberger:2017axi}. First, we write down the energy-momentum tensor for the point particles as
\begin{eqnarray}
T_{pp}^{\mu\nu}=-{2\over \sqrt{g}} {\delta\over \delta g_{\mu\nu}(x)}S_{pp} &=&\sum_{\alpha}\int dx_{\alpha}^{(\mu}p_{\alpha}^{\nu)}\frac{\delta^{d}(x-x_{\alpha})}{\sqrt{g}}e^{\phi} +\int dx_{\alpha}^{(\mu}S_{\alpha}^{\nu)\alpha}\nabla_{\alpha}\bigg[\frac{\delta^{d}(x-x_{\alpha})}{\sqrt{g}}\bigg]\\ &-&\frac{1}{2}\int dx_{\alpha}^{\sigma}H_{\rho\lambda\sigma}g^{\lambda(\mu}S_{\alpha}^{\nu)\rho}e^{-2\phi}\frac{\delta^{d}(x-x_{\alpha})}{\sqrt{g}},\label{eq:emt1}
\end{eqnarray}
where the first line is the result for dilaton gravity, and the second line includes the contribution of the axion. We can also write down the contributions to the energy-momentum tensor from the axion and the dilaton, respectively, as
\begin{equation}
T_{B}^{\mu\nu}=-m_{Pl}^{d-2}e^{-4\phi}g_{\rho\sigma}g_{\tau\lambda}H^{\mu\rho\tau}H^{\nu\sigma\lambda}+\frac{m_{Pl}^{d-2}}{6}e^{-4\phi}H^{2}g^{\mu\nu},\label{eq:emt2}
\end{equation}
and
\begin{equation}
T_{\phi}^{\mu\nu}=(d-2)m_{Pl}^{d-2}(4\partial^{\mu}\phi\partial^{\nu}\phi-2g^{\mu\nu}g^{\rho\sigma}\partial_{\rho}\phi\partial_{\sigma}\phi).\label{eq:emt3}
\end{equation}
Using the equations of motion for the axion and the dilaton, we see that 
\beq
\nabla_{\mu}T_{B}^{\mu\nu}=J_{\sigma\rho}H^{\nu\sigma\rho},
\eeq
and 
\beq
\nabla_{\mu}T_{\phi}^{\mu\nu}=J \del^{\nu} \phi.
\eeq
Now, we obtain the particle equations of motion by integrating $\nabla_{\mu}(T_{pp}^{\mu\nu}+T_{B}^{\mu\nu}+T_{\phi}^{\mu\nu})=0$ with an arbitrary vector $X^{\mu}(x)$, to get 
%After some algebra, we can express the covariant divergence of the
%latter two as
%\begin{equation}
%\nabla_{\mu}(T_{B}^{\mu\nu}+T_{\phi}^{\mu\nu})=-m_{Pl}^{d-2}g_{\rho\sigma}g_{\tau\lambda}H^{\nu\sigma\lambda}\nabla_{\mu}(H^{\mu\rho\tau}e^{-4\phi})+4m_{Pl}^{d-2}(d-2)\bigg(\Box\phi-\frac{1}{6(d-2)}H_{\rho\sigma\lambda}H^{\rho\sigma\lambda}e^{-4\phi}\bigg)\partial^{\nu}\phi.\label{eq:cdemt}
%\end{equation}
%
%Now we can use the exact field equations of motion for 2-form gauge
%field and dilaton, which are given by
%\begin{equation}
%\Box\phi-\frac{e^{-4\phi}}{6(d-2)}H_{\mu\nu\rho}H^{\mu\nu\rho}=-\frac{1}{4(d-2)m_{Pl}^{d-2}}J,\label{eq:eomd}
%\end{equation}
%\begin{equation}
%\nabla_{\lambda}(e^{-4\phi}H^{\mu\nu\lambda})=\frac{1}{m_{Pl}^{d-2}}J^{\mu\nu},\label{eq:eomb}
%\end{equation}
%where we are denoting the RHS as
%\begin{equation}
%J=-\sum_{\alpha}\int ds\bigg(-\dot{x}_{\alpha}^{\mu}p_{\alpha\mu}e^{\phi}-\lambda_{\alpha}S_{\alpha}^{\mu\nu}\dot{x}_{\alpha}^{\sigma}H_{\mu\nu\sigma}e^{-2\phi}\bigg)\frac{\delta^{d}(x-x_{\alpha})}{\sqrt{g}},\label{eq:sd}
%\end{equation}
%\begin{equation}
%J^{\mu\nu}=\sum_{\alpha}\frac{\lambda_{\alpha}}{2}\int ds(S_{\alpha}^{\lambda\mu}\dot{x}_{\alpha}^{\nu}+S_{\alpha}^{\nu\lambda}\dot{x}_{\alpha}^{\mu}+S_{\alpha}^{\mu\nu}\dot{x}_{\alpha}^{\lambda})\nabla_{\lambda}\bigg[e^{-2\phi}\frac{\delta^{d}(x-x_{\alpha})}{\sqrt{g}}\bigg].\label{eq:sb}
%\end{equation}

\begin{align*}
&\int d^{d}x\sqrt{g}X_{\nu}\nabla_{\mu}\bigg(T_{pp}^{\mu\nu}+T_{B}^{\mu\nu}+T_{\phi}^{\mu\nu}\bigg)
= \int d^{d}x\sqrt{g}X_{\nu}\bigg(\nabla_{\mu}T_{pp}^{\mu\nu}-J_{\sigma\rho}H^{\nu\sigma\rho}-J\partial^{\nu}\phi\bigg)\\
= & \sum_{\alpha}\int ds\bigg[\nabla_{\mu}X_{\nu}\bigg(-\frac{1}{2}\dot{x}\cdot\nabla(S_{\alpha}^{\nu\mu})+\frac{1}{2}H_{\lambda\sigma\rho}g^{\lambda[\mu}S_{\alpha}^{\nu]\rho}\dot{x}_{\alpha}^{\sigma}e^{-2\phi}+\dot{x}_{\alpha}^{[\mu}p_{\alpha}^{\nu]}e^{\phi}\bigg)\\
&+X_{\nu}\bigg(\dot{x}_{\alpha}\cdot\nabla(p_{\alpha}^{\nu}e^{\phi}) -\frac{1}{2}R_{\ \mu\lambda\sigma}^{\nu}\dot{x}_{\alpha}^{\mu}S_{\alpha}^{\sigma\lambda}-\partial^{\nu}\phi(\dot{x}_{\alpha}^{\mu}p_{\alpha\mu}e^{\phi}+S_{\alpha}^{\mu\rho}\dot{x}_{\alpha}^{\sigma}H_{\mu\rho\sigma}e^{-2\phi})-\frac{1}{4}\dot{x}_{\alpha}\cdot\nabla(H_{\lambda\sigma\rho}g^{\nu\lambda}S_{\alpha}^{\sigma\rho}e^{-2\phi})\bigg)\bigg].
\end{align*}
Since the vector $X^{\mu}(x)$ is arbitrary, we can equate the coefficients of $X^{\mu}(x)$ and $\nabla_{\mu}X_{\nu}$ to zero, to get
the exact equations for spin and momentum,
\begin{equation}
\frac{dS_{\alpha}^{\mu\nu}}{ds}=p_{\alpha}^{\mu}\dot{x}_{\alpha}^{\nu}e^{\phi}-p_{\alpha}^{\nu}\dot{x}_{\alpha}^{\mu}e^{\phi}-\Gamma_{\sigma\rho}^{\mu}S_{\alpha}^{\rho\nu}\dot{x}_{\alpha}^{\sigma}-\Gamma_{\sigma\rho}^{\nu}S_{\alpha}^{\mu\rho}\dot{x}_{\alpha}^{\sigma}-\frac{1}{2}H_{\lambda\sigma\rho}g^{\lambda\mu}S_{\alpha}^{\nu\rho}\dot{x}_{\alpha}^{\sigma}e^{-2\phi}+\frac{1}{2}H_{\lambda\sigma\rho}g^{\lambda\nu}S_{\alpha}^{\mu\rho}\dot{x}_{\alpha}^{\sigma}e^{-2\phi},
\end{equation}
\begin{align}
\frac{dp_{\alpha}^{\mu}}{ds}= & p_{\alpha\nu}\dot{x}_{\alpha}^{\nu}\partial^{\mu}\phi-p_{\alpha}^{\mu}\dot{x}_{\alpha}^{\sigma}\partial_{\sigma}\phi-\Gamma_{\sigma\rho}^{\mu}\dot{x}_{\alpha}^{\sigma}p_{\alpha}^{\rho}+\frac{1}{2}R_{\ \nu\lambda\sigma}^{\mu}\dot{x}_{\alpha}^{\nu}S_{\alpha}^{\sigma\lambda}e^{-\phi}\nonumber\\
& +\frac{1}{4}\dot{x}_{\alpha}\cdot\nabla(g^{\mu\lambda}H_{\lambda\sigma\rho}S_{\alpha}^{\sigma\rho}e^{-2\phi})e^{-\phi}-\frac{1}{2}S_{\alpha}^{\lambda\rho}\dot{x}_{\alpha}^{\sigma}H_{\lambda\rho\sigma}e^{-3\phi}\partial^{\mu}\phi.
\end{align}
%In addition, directly varying the action with respect to momenta, we have the (on-shell) relation
%\begin{equation}
%e_{\alpha}^{-1}\dot{x}_{\alpha}^{\mu}=p_{\alpha}^{\mu}-\lambda_{\alpha\nu}S_{\alpha}^{\mu\nu}e^{-\phi}.
%\end{equation}
%We require that the primary constraint is preserved over time $\frac{d}{ds}S_{\alpha}^{\mu\nu}p_{\alpha\nu}=0$ to obtain the exact equation of motion by using the equations above. 
%Since we are working to linear order in spin, we can dis
%\begin{equation}
%\frac{dS_{\alpha}^{\mu\nu}}{ds}=p_{\alpha}^{\mu}\dot{x}_{\alpha}^{\nu}e^{\phi}-p_{\alpha}^{\nu}\dot{x}_{\alpha}^{\mu}e^{\phi}-\Gamma_{\sigma\rho}^{\mu}S_{\alpha}^{\rho\nu}\dot{x}_{\alpha}^{\sigma}-\Gamma_{\sigma\rho}^{\nu}S_{\alpha}^{\mu\rho}\dot{x}_{\alpha}^{\sigma}\label{eq:eoms}
%\end{equation}
%\begin{align}
%\frac{dp_{\alpha}^{\mu}}{ds} & =(p_{\alpha\nu}\dot{x}_{\alpha}^{\nu}\partial^{\mu}\phi-p_{\alpha}^{\mu}\dot{x}_{\alpha}^{\sigma}\partial_{\sigma}\phi)-\Gamma_{\sigma\rho}^{\mu}\dot{x}_{\alpha}^{\sigma}p_{\alpha}^{\rho}+\frac{1}{2}R_{\ \nu\lambda\sigma}^{\mu}\dot{x}_{\alpha}^{\nu}S_{\alpha}^{\sigma\lambda}e^{-\phi}.\label{eq:eomp}
%\end{align}
%Substituting into the spin constraint, we get the relation between velocity and momentum as
%\begin{equation}
%\frac{dx_{\alpha}^{\mu}}{ds}=e_{\alpha}p_{\alpha}^{\mu}+e_{\alpha}e^{-\phi}S_{\alpha}^{\mu\nu}\partial_{\nu}\phi.\label{eq:vpreln}
%\end{equation}

We rewrite the dilaton equation of motion Eq.~(\ref{eq:eomd}) as
\begin{equation}
\Box\phi(x)\equiv-\frac{1}{4m_{Pl}^{d-2}(d-2)}\tilde{J}(x),
\end{equation}
where $\Box \equiv\eta^{\mu\nu}\partial_{\mu}\partial_{\nu}$, and we have defined $\tilde{J}(x)$ to include axion and graviton contributions from the LHS of Eq.~(\ref{eq:eomd}). With the dilaton propagator
\begin{equation}
\braket{\phi(k)\phi(-k)}=\frac{i}{4m_{Pl}^{d-2}(d-2)k^2},
\end{equation}
we can formally write the solution as
\begin{equation}
\braket{\phi}(x)=-\frac{1}{2 m^{(d-2)/2}_{Pl} (d-2)^{1/2}}\int_k \frac{e^{-ik\cdot x}}{k^2} \mathcal{A}_{s}(k),
\end{equation} 
thereby defining a canonically normalized dilaton radiation amplitude $\mathcal{A}_{s}(k)=-\frac{1}{2m_{Pl}^{(d-2)/2}(d-2)^{1/2}}\tilde{J}(k)$ for on shell momentum $k^2=0$. 

Choosing the gauge $\del_{\mu} B^{\mu \nu}=0$, we rewrite the equation of motion for the axion, Eq. (\ref{eq:eomh}) as
\begin{equation} \label{eq:eomb}
\Box B^{\mu\nu}(x) \equiv {1\over m_{Pl}^{d-2}} \tilde{ J}^{\mu\nu}(x).
\end{equation}
The gauge condition ensures that the classical axion current  $ {\tilde{ J}}^{\mu\nu}(x)$, so defined, satisfies the conservation equation $\partial_\mu {\tilde{ J}}^{\mu\nu}(x)=0$. In this gauge, the propagator for the axion field is 
\begin{equation}
\langle B_{\mu\nu}(k) B_{\rho\sigma}(-k)\rangle= {i\over 2 m_{Pl}^{d-2} k^2} \left[\eta_{\mu\rho} \eta_{\nu\sigma} - \eta_{\mu\sigma}\eta_{\nu\rho}\right].
\end{equation}
Then, the formal solution to Eq. (\ref{eq:eomb}) is
\begin{equation}
\braket{B^{\mu\nu}}(x)=-\frac{1}{m^{d-2}_{Pl}}\int_k \frac{e^{-ik\cdot x}}{k^2} \tilde{J}^{\mu\nu}(k),
\end{equation} 
which defines an axion radiation amplitude $\mathcal{A}_B=\frac{1}{m_{Pl}^{(d-2)/2}}a^*_{\mu\nu}\tilde{J}^{\mu\nu}(k)$ for on-shell momentum $k^2=0$.

Finally, we get to the gravitational field. We expand the metric perturbatively about flat space $g_{\mu \nu}=\eta_{\mu \nu}+h_{\mu \nu}$. We choose the de-Donder gauge $\del_{\mu} h^{\mu \nu}=\half \del_{\mu}h^{\lambda}_{\lambda}$, in which we have the propagator
\begin{equation}
\langle h_{\mu\nu}(k) h_{\rho\sigma}(-k)\rangle= {i\over 2 m_{Pl}^{d-2} k^2} \left[\eta_{\mu\rho} \eta_{\nu\sigma} + \eta_{\mu\sigma}\eta_{\nu\rho}-\frac{2}{d-2} \eta_{\mu\nu} \eta_{\rho\sigma}\right],
\end{equation}
and the solution for $h_{\mu \nu}(x)$ can be written formally as
\beq \label{eq:hsol}
\braket{h_{\mu \nu}}(x) =  {1\over 2m_{Pl}^{d-2}  } \int {e^{-ik\cdot x}\over k^2} \left\{{\tilde T}_{\mu \nu}(k)-{1 \over {d-2}} \eta_{\mu \nu} {\tilde T}^{\sigma}_{\sigma}\right\}.
\eeq
Here, ${\tilde T}_{\mu \nu}(k)$ is the energy-momentum pseudo tensor that includes contributions from the sources as well as all the fields. It is (non-covariantly) conserved, $\del_{\mu} {\tilde T}^{\mu \nu}=0$, and coordinate dependent, hence non-unique. However, quantities such as energy-momentum or angular momentum can be defined as suitable integrals of ${\tilde T}^{\mu \nu}$ over spacetime.  As in \cite{Goldberger:2004jt}, we compute the background field gauge effective action~\cite{BgGauge}, expressed as
\beq
\Gamma (h,\phi,B)=-\half \int d^d x \, {\tilde T}^{\mu \nu}h_{\mu \nu},
\eeq
so that the energy-momentum pseudotensor is related to the coefficient of the graviton 1-point function. It is also directly related to the graviton amplitude by $\mathcal{A}_g=-\frac{1}{2m_{Pl}^{(d-2)/2}}\epsilon^*_{\mu\nu}{\tilde T}^{\mu \nu}(k)$.

Observables at infinity are obtained in a manner similar to Eq. (\ref{eq:pw}) by replacing the gluon source current and polarization vectors with the corresponding gravity ones. Hence, in the following, we solve for these sources ${\tilde T}^{\mu \nu}(k)$,  $\tilde{J}^{\mu\nu}(k)$, and $\tilde{J}(k)$, in a perturbative expansion in powers of $\eta$, defined in Eq. (\ref{eq:ccdc}), and to linear order in spin.

\subsection{Leading order results}
\label{sec:LO}
We now find the leading order solutions to the fields and the particle equations of motion. At leading order in perturbation, the axion field only gets a contribution  from Fig.~\ref{fig:axion1pt}(a)
\beq
%B^{\mu\nu}(x)=- i \sum_{\alpha}\frac{\lambda_{\alpha}}{2m_{Pl}^{d-2}} (S_{\alpha}^{\mu\lambda}\dot{x}_{\alpha}^{\nu}-S_{\alpha}^{\nu\lambda}\dot{x}_{\alpha}^{\mu})\int_k  \left(2\pi\right) \delta(k\cdot p_{\alpha})  \frac{e^{-i k \cdot (x-b_\alpha)}}{k^2} k_\lambda.
\braket{B^{\mu\nu}}(x)\bigr\vert_{\mathcal{O}(\eta^2)}= \frac{i}{4m_{Pl}^{d-2}} \sum_{\alpha} \int {d^d\ell\over (2\pi)^d} {e^{-i\ell\cdot (x- x_\alpha)}\over\ell^2}\Big[(S_{\alpha}\wedge  \ell)^{\mu}p_{\alpha}^{\nu} -(S_{\alpha}\wedge \ell)^{\nu}p_{\alpha}^{\mu}-S_\alpha^{\mu\nu}(\ell\cdot p_\alpha)\Big]
\eeq
for general time-dependent dynamical variables $p^\mu_\alpha$ and $S^{\mu\nu}_\alpha$ that satisfy the equations of motion.

The leading order metric perturbation contains a spinning and a non-spinning contribution sourced by Fig.~\ref{fig:axion1pt}(a) and (b) respectively. Their sum is given by 
\begin{equation}
\braket{h^{\mu\nu}}(x)\bigr\vert_{\mathcal{O}(\eta^2)}=\frac{1}{2m_{Pl}^{d-2}}\sum_{\alpha}\int  {d^d\ell\over (2\pi)^d} {e^{-i\ell\cdot (x- x_\alpha)}\over\ell^2}\bigg[p_{\alpha}^{\mu}p_{\alpha}^{\nu}-\frac{p^2_\alpha}{d-2}\eta_{\mu \nu}-\frac{i}{2} \Big\{ p_{\alpha}^{\mu}(S_{\alpha}\wedge \ell)^{\nu}+p_{\alpha}^{\nu}(S_{\alpha}\wedge \ell)^{\mu}\Big\}\bigg].
\end{equation}

The leading order dilaton solution has no spin dependent contribution, so it is the same as in the non-spinning case \cite{Goldberger:2016iau},
\begin{equation}
\braket{\phi}(x)\bigr\vert_{\mathcal{O}(\eta^2)}=\frac{1}{4m_{Pl}^{d-2}(d-2)}\sum_\alpha \int {d^d\ell\over (2\pi)^d} {e^{-i\ell\cdot (x- x_\alpha)}\over\ell^2}p_\alpha^2.
\end{equation}

%Just as we did in the Yang-Mills section, we can easily generalize to the case where at initially the particles follow time-dependent orbits (that are solutions to the equations of motion) $x^\mu_\alpha(\tau_{\alpha})=x^\mu_\alpha$, $p^\mu_\alpha(\tau_{\alpha})=p^\mu_\alpha$, $ c^a_\alpha(\tau_{\alpha})=c^a_\alpha $. In this case, 
Next, we calculate the corrections in spin and momentum that these leading order fields induce. First, let us write the equations of motion for the particle up to linear order in spin,
\begin{equation}
\frac{d}{ds}S_{\alpha}^{\mu\nu}\bigr\vert_{\mathcal{O}(S^0)+\mathcal{O}(S^1)}=p_{\alpha}^{\mu}v_{\alpha}^{\nu}e^{\phi}-p_{\alpha}^{\nu}v_{\alpha}^{\mu}e^{\phi}-\Gamma_{\sigma\rho}^{\mu}S_{\alpha}^{\rho\nu}v_{\alpha}^{\sigma}-\Gamma_{\sigma\rho}^{\nu}S_{\alpha}^{\mu\rho}v_{\alpha}^{\sigma},\label{eq:eoms}
\end{equation}
\begin{align}
\frac{d}{ds}p_{\alpha}^{\mu}\bigr\vert_{\mathcal{O}(S^0)+\mathcal{O}(S^1)} & =p_{\alpha\nu}v_{\alpha}^{\nu}\partial^{\mu}\phi-p_{\alpha}^{\mu}v_{\alpha}^{\sigma}\partial_{\sigma}\phi-\Gamma_{\sigma\rho}^{\mu}v_{\alpha}^{\sigma}p_{\alpha}^{\rho}+\frac{1}{2}R_{\ \nu\lambda\sigma}^{\mu}v_{\alpha}^{\nu}S_{\alpha}^{\sigma\lambda}e^{-\phi}.\label{eq:eomp}
\end{align}
Inserting the leading order fields into the above equations give the leading order changes in momenta and spin. First, for the momenta, we have
\begin{align}
\frac{d}{d\tau_{\alpha}}p_{\alpha}^{\mu}\bigr\vert_{\mathcal{O}(\eta^2,S^0)} & =\frac{i}{4m_{Pl}^{d-2}}\sum_{\beta\neq\alpha}\int d\tau_\beta \frac{d^d\ell}{(2\pi)^d} \frac{e^{-i\ell\cdot x_{\alpha\beta}}}{\ell^{2}}\bigg[-\frac{m_{\beta}^{2}}{d-2}(\ell\cdot p_{\alpha})p_{\alpha}^{\nu}+2(p_{\alpha}\cdot p_{\beta})(\ell\cdot p_{\alpha})p_{\beta}^{\nu}-(p_{\alpha}\cdot p_{\beta})^{2}\ell^{\nu}\bigg],
\end{align}
\begin{align}
\frac{d}{d\tau_{\alpha}}p_{\alpha}^{\mu}\bigr\vert_{\mathcal{O}(\eta^2,S^1)} & =\frac{1}{4m_{Pl}^{d-2}}\sum_{\beta\neq\alpha}\int d\tau_\beta \frac{d^d\ell}{(2\pi)^d} \frac{e^{-i\ell\cdot x_{\alpha\beta}}}{\ell^{2}}\bigg[\frac{p_{\beta}^{2}}{d-2}(\ell\cdot p_{\alpha})(S_{\alpha}\wedge \ell)^{\mu}\nonumber \\
& +[(\ell\wedge p_{\beta})_{\alpha}-(\ell\wedge p_{\alpha})_\beta][(\ell\cdot p_{\alpha})p_{\beta}^{\mu}-(p_{\alpha}\cdot p_{\beta})\ell^{\mu}]+(\ell\cdot p_{\alpha})(p_{\alpha}\cdot p_{\beta})(S_{\beta}\wedge \ell)^{\mu}\bigg].\label{eq:peom}
\end{align}
From this, we have the position equations of motion
\begin{align}
v_{\alpha}^{\mu} & =p_{\alpha}^{\mu}+w_{\alpha}^{\mu}\bigr\vert_{\mathcal{O}(\eta^2,S^1)},\label{eq:xeom}
\end{align}
\begin{equation}
w_{\alpha}^{\mu}\bigr\vert_{\mathcal{O}(\eta^2,S^1)}\equiv-\frac{i}{4m_{Pl}^{d-2}(d-2)}\sum_{\beta\neq\alpha}\int d\tau_\beta \frac{d^d\ell}{(2\pi)^d} \frac{e^{-i\ell\cdot x_{\alpha\beta}}}{\ell^{2}}p_{\beta}^{2}(S_{\alpha}\wedge \ell)^{\mu}.\label{eq:xeom2}
\end{equation}
Similarly, the spin equation of motion is given by
\begin{align}
\frac{d}{d\tau_{\alpha}}S_{\alpha}^{\mu\nu}\bigr\vert_{\mathcal{O}(\eta^2,S^1)} & =p_{\alpha}^{\mu}w_{\alpha}^{\nu}\bigr\vert_{\mathcal{O}(\eta^2,S^1)}-p_{\alpha}^{\nu}w_{\alpha}^{\mu}\bigr\vert_{\mathcal{O}(\eta^2,S^1)}\nonumber \\
& +\frac{i}{4m_{Pl}^{d-2}}\sum_{\beta\neq\alpha}\int d\tau_\beta \frac{d^d\ell}{(2\pi)^d} \frac{e^{-i\ell\cdot x_{\alpha\beta}}}{\ell^{2}}\bigg[(\ell\cdot p_{\alpha})\Big\{p_{\beta}^{\nu}(S_{\alpha}\wedge p_{\beta})^{\mu}-p_{\beta}^{\mu}(S_{\alpha}\wedge p_{\beta})^{\nu}\Big\}\nonumber \\
&+(p_{\alpha}\cdot p_{\beta})\Big\{p_{\beta}^{\nu}(S_{\alpha}\wedge \ell)^{\mu}-p_{\beta}^{\mu}(S_{\alpha}\wedge \ell)^{\nu}+\ell^{\mu}(S_{\alpha}\wedge p_{\beta})^{\nu}-\ell^{\nu}(S_{\alpha}\wedge p_{\beta})^{\mu}\Big\}\nonumber
\\
& +\frac{p_{\beta}^{2}}{d-2}\Big\{p_{\alpha}^{\mu}(S_{\alpha}\wedge \ell)^{\nu}-p_{\alpha}^{\nu}(S_{\alpha}\wedge \ell)^{\mu}-\ell^{\mu}(S_{\alpha}\wedge p_{\alpha})^{\nu}+\ell^{\nu}(S_{\alpha}\wedge p_{\alpha})^{\mu}-2(\ell\cdot p_{\alpha})S_{\alpha}^{\mu\nu}\Big\}\bigg].\label{eq:seom}
\end{align}

\subsection{Axion radiation}
\label{sec:axirad}
\begin{figure}
	\centering
	\includegraphics[scale=0.59]{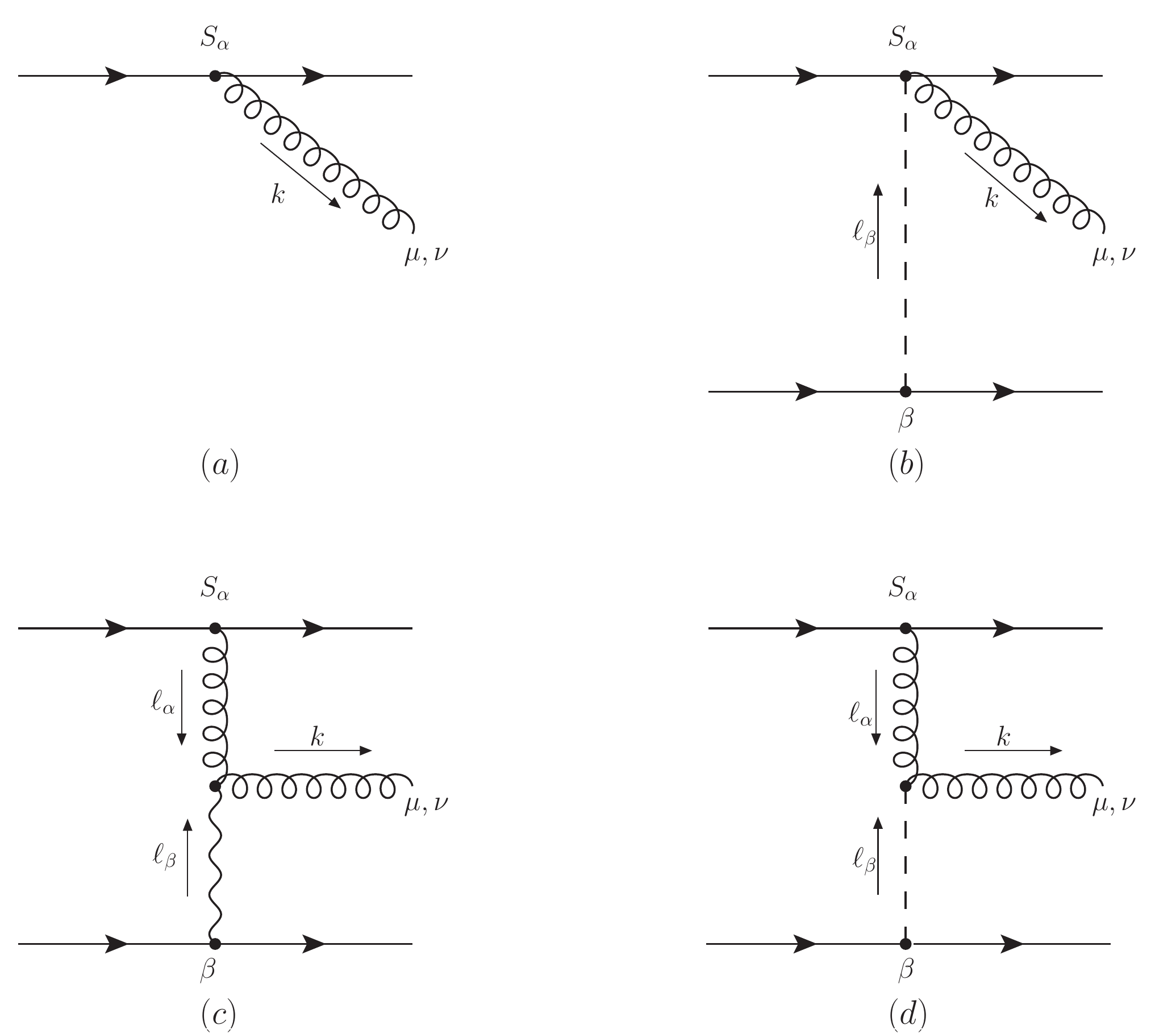}
	\caption{\label{fig:axion1pt} Feynman diagrams for the perturbative expansion of the axion source current $\tilde{ J}^{\mu\nu}(k)$ up to $\mathcal{O}(\eta^2)$, with a single spin insertion. Here, wavy lines, curvy lines and dashed lines respectively represent gravitons, axions and dilatons. Diagrams $(a)-(b)$ represent axion radiation coming directly off the worldline. Diagrams $(c)-(d)$ correspond to axion radiation from bulk dilaton and graviton vertices.}    
\end{figure}
The leading order radiation has contributions from Figs.~\ref{fig:axion1pt}(a)-(d). The contribution from Fig.~\ref{fig:axion1pt}(a) is due to deflections in the position and spin of the particles, induced by the leading order fields. It is given by
\begin{subequations}
\begin{align}
\tilde{ J}^{\mu\nu}(k)\bigr\vert_{\text{Fig.~\ref{fig:axion1pt}(a)};\mathcal{O}(\eta^2,S^1)}=\frac{i}{4}\sum_{\alpha}&\int d\tau_\alpha e^{ik\cdot x_\alpha}{k_{\lambda}\over k\cdot v_\alpha}\bigg[-\frac{k\cdot \dot{v}_{\alpha}}{k\cdot v_\alpha}S_{\alpha}^{\lambda\mu}v_{\alpha}^{\nu}+ S_{\alpha}^{\lambda\mu}\dot{v}_\alpha^\nu\label{line12}\\
&+\dot{S}_{\alpha}^{\lambda\mu}v_{\alpha}^{\nu}+\text{cyclic permutations $(\mu,\nu,\lambda)$}\bigg]\Biggr\vert_{\mathcal{O}(\eta^2,S^1)}.\label{line22}
\end{align}
\end{subequations}
Substituting the corresponding changes in spin and momenta derived in the previous subsection gives
\begin{align}
\text{(\ref{line12})}	&=-\frac{i}{16m_{Pl}^{d-2}}\sum_{{\alpha,\beta\atop \alpha\neq\beta}}\int d\mu_{\alpha\beta}(k)\ell_{\alpha}^{2}\bigg[-\frac{(p_{\alpha}\cdot p_{\beta})^{2}(k\cdot\ell_{\beta})}{(k\cdot p_{\alpha})^{2}}(S_{\alpha}\wedge k)^{\mu}p_{\alpha}^{\nu}\nonumber\\
& +\frac{p_{\alpha}\cdot p_{\beta}}{k\cdot p_{\alpha}}\Big\{(p_{\alpha}\cdot p_{\beta})\ell_{\beta}^{\nu}-2(k\cdot p_{\alpha})p_{\beta}^{\nu}+2(k\cdot p_{\beta})p_{\alpha}^{\nu}\Big\}(S_{\alpha}\wedge k)^{\mu}-(\mu\leftrightarrow\nu)\bigg],
\end{align}
\begin{align}
\text{(\ref{line22})}=  -\frac{i}{16m_{Pl}^{d-2}}&\sum_{{\alpha,\beta\atop \alpha\neq\beta}}\int d\mu_{\alpha\beta}(k)\ell_{\alpha}^{2}\Bigg[\frac{p_{\alpha}\cdot p_{\beta}}{k\cdot p_{\alpha}}\Big\{(k\cdot\ell_{\beta})(S_{\alpha}\wedge p_{\beta})^{\mu}p_{\beta}^{\nu}-(k\cdot p_{\beta})(S_{\alpha}\wedge\ell_{\beta})^{\mu}p_{\beta}^{\nu}-(k\wedge\ell_{\beta})_{\alpha}p_{\alpha}^{\mu}p_{\beta}^{\nu}\nonumber\\
& +(k\wedge p_{\beta})_{\alpha}p_{\alpha}^{\mu}\ell_{\beta}^{\nu}\Big\}+(p_{\alpha}\cdot p_{\beta})\Big\{(S_{\alpha}\wedge\ell_{\beta})^{\mu}p_{\beta}^{\nu}-(S_{\alpha}\wedge p_{\beta})^{\mu}\ell_{\beta}^{\nu}\Big\}+(k\cdot p_{\alpha})(S_{\alpha}\wedge p_{\beta})^{\mu}p_{\beta}^{\nu}\nonumber\\
&-(k\cdot p_{\beta})(S_{\alpha}\wedge p_{\beta})^{\mu}p_{\alpha}^{\nu}-(k\wedge p_{\beta})_{\alpha}p_{\alpha}^{\mu}p_{\beta}^{\nu}+\frac{p_{\beta}^{2}}{d-2}\Big\{2(S_{\alpha}\wedge k)^{\mu}-(k\cdot p_{\alpha})S_{\alpha}^{\mu\nu}\Big\}-(\mu\leftrightarrow\nu)\bigg].
\end{align}

The other contributions to the axion amplitude, at this order in perturbation, come from diagrams with no deflections in the trajectories of the particles. Fig.~\ref{fig:axion1pt}(b), with an intermediate dilaton, corresponds to
\begin{equation}
\tilde{J}^{\mu\nu}(k)\bigr\vert_{\text{Fig.~\ref{fig:axion1pt}(b)};\mathcal{O}(\eta^{2},S^{1})}=-\frac{i}{16m_{Pl}^{d-2}(d-2)}\sum_{{\alpha,\beta\atop \alpha\neq\beta}}\int d\mu_{\alpha\beta}(k)\ell_{\alpha}^{2}p_{\beta}^{2}\bigg[(k\cdot p_{\alpha})S_{\alpha}^{\mu\nu}-2(S_{\alpha}\wedge k)^{\mu}p_{\alpha}^{\nu}-(\mu\leftrightarrow\nu)\bigg].
\end{equation}

The two 3-point vertex diagrams in Figs.~\ref{fig:axion1pt}(c),(d) contribute
\begin{align}
\tilde{J}^{\mu\nu}(k)\bigr\vert_{\text{Fig.~\ref{fig:axion1pt}(c)};\mathcal{O}(\eta^{2},S^{1})}= & -\frac{i}{4m_{Pl}^{d-2}(d-2)}\sum_{{\alpha,\beta\atop \alpha\neq\beta}}\int d\mu_{\alpha\beta}(k)\bigg[p_{\beta}^{2}\Big\{(k\cdot\ell_{\alpha})(S_{\alpha}\wedge\ell_{\alpha})^{\mu}p_{\alpha}^{\nu}\nonumber\\
& -(k\cdot p_{\alpha})(S_{\alpha}\wedge\ell_{\alpha})^{\mu}\ell_{\alpha}^{\nu}+(k\wedge\ell_{\alpha})_{\alpha}p_{\alpha}^{\mu}\ell_{\alpha}^{\nu}\Big\}-(\mu\leftrightarrow\nu)\bigg],
\end{align}
\begin{align}
\tilde{J}^{\mu\nu}(k)&\bigr\vert_{\text{Fig.~\ref{fig:axion1pt}(d)};\mathcal{O}(\eta^{2},S^{1})}= -\frac{i}{8m_{Pl}^{d-2}}\sum_{{\alpha,\beta\atop \alpha\neq\beta}}\int d\mu_{\alpha\beta}(k)\bigg[(p_{\alpha}\cdot p_{\beta})\Big\{(k\cdot\ell_{\alpha})(S_{\alpha}\wedge\ell_{\alpha})^{\mu}p_{\beta}^{\nu}-(k\cdot p_{\beta})(S_{\alpha}\wedge\ell_{\alpha})^{\mu}\ell_{\alpha}^{\nu}\nonumber\\
& -(k\wedge\ell_{\alpha})_{\alpha}l_{\alpha}^{\mu}p_{\beta}^{\nu}\Big\}+(k\cdot p_{\beta})^{2}(S_{\alpha}\wedge\ell_{\alpha})^{\mu}p_{\alpha}^{\nu}-(k\cdot p_{\alpha})(k\cdot p_{\beta})(S_{\alpha}\wedge\ell_{\alpha})^{\mu}p_{\beta}^{\nu}+(k\cdot p_{\beta})(k\wedge\ell_{\alpha})_{\alpha}p_{\alpha}^{\mu}p_{\beta}^{\nu}\nonumber\\
& +(k\cdot p_{\beta})(\ell_{\alpha}\wedge p_{\beta})_\alpha\ell_{\alpha}^{\mu}p_{\alpha}^{\nu}-(k\cdot p_{\alpha})(\ell_{\alpha}\wedge p_{\beta})_{\alpha}\ell_{\alpha}^{\mu}p_{\beta}^{\nu}+(k\cdot\ell_{\alpha})(\ell_{\alpha}\wedge p_{\beta})_{\alpha}p_{\alpha}^{\mu}p_{\beta}^{\nu}\nonumber\\
& -\frac{2p_{\beta}^{2}}{d-2}\Big\{(k\cdot\ell_{\alpha})(S_{\alpha}\wedge\ell_{\alpha})^{\mu}p_{\alpha}^{\nu}-(k\cdot p_{\alpha})(S_{\alpha}\wedge\ell_{\alpha})^{\mu}\ell_{\alpha}^{\nu}+(k\wedge\ell_{\alpha})_{\alpha}p_{\alpha}^{\mu}\ell_{\alpha}^{\nu}\Big\}-(\mu\leftrightarrow\nu)\bigg].
\end{align}

%The contribution from the 3-point vertex with dilaton
%\bea
%{B}^{\mu\nu}(k)&=&i\sum_{\alpha,\beta\atop \alpha\neq\beta}\int_{l_{\alpha},l_{\beta}}\frac{\mu_{\alpha,\beta}(k)}{16m_{Pl}^{3(d-2)/ 2}} \left(-p_{\alpha }^{\nu }k \cdot  l_{\alpha } \left(S_{\alpha }^{\mu }\cdot l_{\alpha } \right)+l_{\alpha }^{\nu }k \cdot  p_{\alpha } \left(S_{\alpha }^{\mu }\cdot l_{\alpha } \right)+p_{\alpha }^{\mu }k \cdot  l_{\alpha } \left(S_{\alpha }^{\nu }\cdot l_{\alpha} \right)-l_{\alpha }^{\mu }k \cdot  p_{\alpha } \left(S_{\alpha }^{\nu }\cdot l_{\alpha} \right)+k\cdot S_{\alpha }\cdot l_{\alpha } \left(l_{\alpha }^{\mu } p_{\alpha }^{\nu }-l_{\alpha }^{\nu } p_{\alpha }^{\mu }\right) \right) \nonumber \\
%\eea
We notice that the Yang-Mills amplitude has no explicit dependence on the space-time dimension, whereas some of the contributions to the axion amplitude above do. This means that terms involving the dimension of space-time should cancel with each other, giving the total axion amplitude
\begin{align}
\mathcal{A}_{B}(k)\bigr\vert&_{\mathcal{O}(\eta^{3},S^{1})}  =-\frac{ia^*_{\mu\nu}}{16m_{Pl}^{3(d-2)/2}}\sum_{{\alpha,\beta\atop \alpha\neq\beta}}\int d\mu_{\alpha\beta}(k)\bigg[-\frac{(p_{\alpha}\cdot p_{\beta})^{2}(k\cdot\ell_{\beta})\ell_{\alpha}^{2}}{(k\cdot p_{\alpha})^{2}}(S_{\alpha}\wedge k)^{\mu}p_{\alpha}^{\nu}+\frac{(p_{\alpha}\cdot p_{\beta})\ell_{\alpha}^{2}}{k\cdot p_{\alpha}}\bigg\{\Big((p_{\alpha}\cdot p_{\beta})\ell_{\beta}^{\nu}\nonumber\\
& -2(k\cdot p_{\alpha})p_{\beta}^{\nu}+2(k\cdot p_{\beta})p_{\alpha}^{\nu}\Big)(S_{\alpha}\wedge k)^{\mu}+(k\cdot\ell_{\beta})(S_{\alpha}\wedge p_{\beta})^{\mu}p_{\beta}^{\nu}-(k\cdot p_{\beta})(S_{\alpha}\wedge\ell_{\beta})^{\mu}p_{\beta}^{\nu}-(k\wedge\ell_{\beta})_{\alpha}p_{\alpha}^{\mu}p_{\beta}^{\nu}\nonumber\\
& +(k\wedge p_{\beta})_{\alpha}p_{\alpha}^{\mu}\ell_{\beta}^{\nu}\bigg\}+(p_{\alpha}\cdot p_{\beta})\Big\{\ell_{\alpha}^{2}(S_{\alpha}\wedge\ell_{\beta})^{\mu}p_{\beta}^{\nu}-\ell_{\alpha}^{2}(S_{\alpha}\wedge p_{\beta})^{\mu}\ell_{\beta}^{\nu}+2(k\cdot\ell_{\alpha})(S_{\alpha}\wedge\ell_{\alpha})^{\mu}p_{\beta}^{\nu}\nonumber\\
& -2(k\cdot p_{\beta})(S_{\alpha}\wedge\ell_{\alpha})^{\mu}\ell_{\alpha}^{\nu}-2(k\wedge\ell_{\alpha})_{\alpha}l_{\alpha}^{\mu}p_{\beta}^{\nu}\Big\}+\ell_{\alpha }^2(k\cdot p_{\alpha})(S_{\alpha}\wedge p_{\beta})^{\mu}p_{\beta}^{\nu}-\ell_{\alpha }^2(k\cdot p_{\beta})(S_{\alpha}\wedge p_{\beta})^{\mu}p_{\alpha}^{\nu}\nonumber\\
& -2(k\cdot p_{\alpha})(k\cdot p_{\beta})(S_{\alpha}\wedge\ell_{\alpha})^{\mu}p_{\beta}^{\nu}+2(k\cdot p_{\beta})^{2}(S_{\alpha}\wedge\ell_{\alpha})^{\mu}p_{\alpha}^{\nu}-\ell_{\alpha }^2(k\wedge p_{\beta})_{\alpha}p_{\alpha}^{\mu}p_{\beta}^{\nu}+2(k\cdot p_{\beta})(k\wedge\ell_{\alpha})_{\alpha}p_{\alpha}^{\mu}p_{\beta}^{\nu}\nonumber\\
& +2(k\cdot p_{\beta})(\ell_{\alpha}\wedge p_{\beta})_\alpha\ell_{\alpha}^{\mu}p_{\alpha}^{\nu}-2(k\cdot p_{\alpha})(\ell_{\alpha}\wedge p_{\beta})_{\alpha}\ell_{\alpha}^{\mu}p_{\beta}^{\nu}+2(k\cdot\ell_{\alpha})(\ell_{\alpha}\wedge p_{\beta})_{\alpha}p_{\alpha}^{\mu}p_{\beta}^{\nu}-(\mu\leftrightarrow\nu)\bigg].
\end{align}

Each diagram satisfies the Ward identity $k_\mu \tilde{J}^{\mu\nu}(k)=0$ and so does the total amplitude. The total amplitude is in agreement with the calculation in string frame metric, presented in \cite{Goldberger:2017axi}. We find this matches the double copy result Eq.~(\ref{eq:bfielddc}).

\subsection{Graviton radiation}
\label{sec:gravrad}
\begin{figure}
	\centering
	\includegraphics[scale=.59]{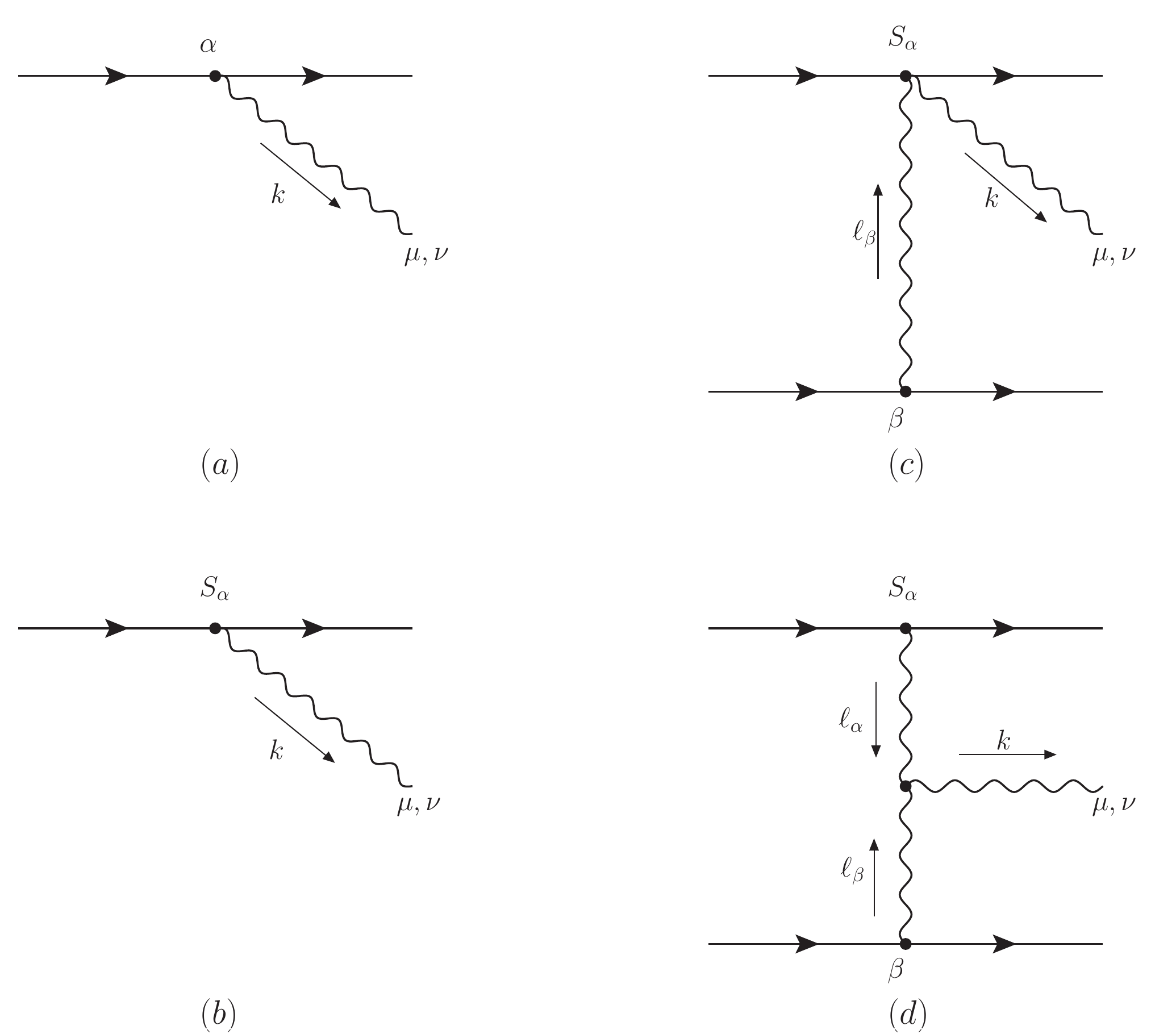}
		\caption{\label{fig:graviton1pt}Feynman diagrams that contribute to the energy-momentum pseudotensor $\tilde{T}^{\mu\nu}(k)$ at $\mathcal{O}(\eta^2)$, with a single spin insertion. Diagram $(a)$ represents graviton radiation from corrections to the spin-independent piece due to the equations of motion.  Diagrams $(b)-(d)$ correspond to corrections at linear order in spin.}    
\end{figure}
Similarly, at the next to leading order in perturbation, the energy momentum pseudotensor receives contributions from Fig.~\ref{fig:graviton1pt}(a)-(d).
The contribution from Figs.~\ref{fig:graviton1pt}(a) and (b) come from deflections to the particle spin and trajectory due to the leading order fields. This comes out to be
\begin{subequations}
\begin{align}
\tilde{T}^{\mu\nu}(k)\bigr\vert_{\text{Fig.~\ref{fig:graviton1pt}(a)+(b)};\mathcal{O}(\eta^{2},S^1)}   =\frac{1}{2}\sum_{\alpha}\int d\tau_\alpha e^{ik\cdot x_\alpha}\bigg[&\frac{-i}{k\cdot v_\alpha}\bigg\{\frac{k\cdot \dot{v}_\alpha}{k\cdot v_\alpha}v_\alpha^\mu p_\alpha^\nu-\dot{v}_\alpha^\mu p_\alpha^\nu -v^\mu \dot{p}^\nu_\alpha\bigg\}\label{eq:gravtraj}\\
& -\frac{k_\rho}{k\cdot v_\alpha}\bigg\{\frac{k\cdot \dot{v}_\alpha}{k\cdot v_\alpha}v_\alpha^\mu S^{\nu\rho}_\alpha-\dot{v}^\mu_{\alpha} S^{\nu\rho}_\alpha\bigg\}+(\mu\leftrightarrow \nu)\bigg]\Biggr\vert_{\mathcal{O}(\eta^{2},S^1)}.\label{eq:gravsp}
\end{align}
\end{subequations}
These are explicitly given by
\begin{align}
\text{(\ref{eq:gravtraj})} & =\frac{i}{4m_{Pl}^{d-2}}\sum_{\alpha,\beta\atop\alpha\neq\beta}\int d\mu_{\alpha\beta}(k)\ell_{\alpha}^{2}\bigg[\frac{(p_{\alpha}\cdot p_{\beta})(k\cdot \ell_{\beta})}{2(k\cdot p_{\alpha})^{2}}\Big\{(\ell_{\beta}\wedge p_{\beta})_{\alpha}-(\ell_{\beta}\wedge p_{\alpha})_{\beta}\Big\} p_{\alpha}^{\mu}p_{\alpha}^{\nu}\nonumber \\
& -\frac{p_{\alpha}\cdot p_{\beta}}{k\cdot p_{\alpha}}\bigg\{\Big((\ell_{\beta}\wedge p_{\beta})_{\alpha}-(\ell_{\beta}\wedge p_{\alpha})_{\beta}\Big)p_{\alpha}^{\mu}\ell_{\beta}^{\nu}+\frac{1}{2}(\ell_{\alpha}\wedge \ell_{\beta})_{\beta}p_{\alpha}^{\mu}p_{\alpha}^{\nu}\bigg\}-\frac{k\cdot p_{\beta}}{2(k\cdot p_{\alpha})}\Big\{(\ell_{\beta}\wedge p_{\beta})_{\alpha}-(\ell_{\beta}\wedge p_{\alpha})_{\beta}\Big\} p_{\alpha}^{\mu}p_{\alpha}^{\nu}\nonumber \\
& +(p_{\alpha}\cdot p_{\beta})(S_{\beta}\wedge \ell_{\beta})^{\mu}p_{\alpha}^{\nu}+\Big\{(\ell_{\beta}\wedge p_{\beta})_{\alpha}-(\ell_{\beta}\wedge p_{\alpha})_{\beta}\Big\} p_{\alpha}^{\mu}p_{\beta}^{\nu}+\frac{m_{\beta}^{2}}{2(d-2)}(S_{\alpha}\wedge \ell_{\beta})^{\mu}p_{\alpha}^{\nu}+(\mu\leftrightarrow\nu)\bigg],\label{eq:gtraj}
\end{align}
\begin{align}
\text{(\ref{eq:gravsp})} & =\frac{i}{4m_{Pl}^{d-2}}\sum_{\alpha,\beta\atop\alpha\neq\beta}\int d\mu_{\alpha\beta}(k)\ell_{\alpha}^{2}\bigg[\frac{(p_{\alpha}\cdot p_{\beta})^{2}(k\cdot \ell_{\beta})}{2(k\cdot p_{\alpha})^{2}}(S_{\alpha}\wedge k)^{\mu}p_{\alpha}^{\nu}-\frac{(p_{\alpha}\cdot p_{\beta})^{2}}{2(k\cdot p_{\alpha})}(S_{\alpha}\wedge k)^{\mu}\ell_{\beta}^{\nu}\nonumber \\
& -\frac{p_{\alpha}\cdot p_{\beta}}{2(k\cdot p_{\alpha})}\Big\{(k\cdot p_{\beta})\Big((S_{\alpha}\wedge k)^{\mu}+(S_{\alpha}\wedge \ell_{\alpha})^{\mu}\Big)p_{\alpha}^{\nu}+(k\cdot \ell_{\beta})(S_{\alpha}\wedge p_{\beta})^{\mu}p_{\alpha}^{\nu}+(\ell_{\alpha}\wedge \ell_{\beta})_{\alpha}p_{\alpha}^{\mu}p_{\beta}^{\nu}-(k\wedge p_{\beta})_{\alpha}p_{\alpha}^{\mu}\ell_{\beta}^{\nu}\Big\}\nonumber \\
& +(p_{\alpha}\cdot p_{\beta})(S_{\alpha}\wedge k)^{\mu}p_{\beta}^{\nu}+\frac{1}{2}(k\cdot p_{\beta})(S_{\alpha}\wedge p_{\beta})^{\mu}p_{\alpha}^{\nu}-\frac{1}{2}(k\wedge p_{\beta})_{\alpha}p_{\alpha}^{\mu}p_{\beta}^{\nu}-\frac{m_{\beta}^{2}}{d-2}(S_{\alpha}\wedge k)^{\mu}p_{\alpha}^{\nu}+(\mu\leftrightarrow\nu)\bigg].\label{eq:gsp}
\end{align}

There are two contributions from emission off bulk vertices with the particles not suffering any deflections. The first of these is from Fig.~\ref{fig:graviton1pt}(c) with an intermediate
graviton. This contributes
\begin{align}
\tilde{T}^{\mu\nu}(k)\bigr\vert_{\text{Fig.~\ref{fig:graviton1pt}(c)};\mathcal{O}(\eta^{2},S^1)} & =\frac{i}{4m_{Pl}^{d-2}}\sum_{\alpha,\beta\atop\alpha\neq\beta}\int d\mu_{\alpha\beta}(k)\frac{\ell_{\alpha}^{2}}{2}\Big[(p_{\alpha}\cdot p_{\beta})\Big\{(S_{\alpha}\wedge p_{\beta})^{\mu}\ell_{\beta}^{\nu}-(S_{\alpha}\wedge \ell_{\beta})^{\mu}p_{\beta}^{\nu}\Big\}\nonumber \\
& -(k\cdot p_{\alpha})(S_{\alpha}\wedge p_{\beta})^{\mu}p_{\beta}^{\nu}+\frac{m_{\beta}^{2}}{d-2}(S_{\alpha}\wedge \ell_{\beta})^{\mu}p_{\alpha}^{\nu}+(\mu\leftrightarrow\nu)\Big].\label{eq:gstat}
\end{align}

The final contribution is from the graviton triple vertex diagram in Fig.~\ref{fig:graviton1pt}(d). As in \cite{Goldberger:2016iau}, in computing this contribution, we use the background field gauge 3-point vertex, written, for example, in \cite{Donoghue:1994dn, Goldberger:2004jt}. This gives
\begin{align}
\tilde{T}^{\mu\nu}(k)\bigr\vert_{\text{Fig.~\ref{fig:graviton1pt}(d)};\mathcal{O}(\eta^{2},S^1)} & =\frac{i}{4m_{Pl}^{d-2}}\sum_{\alpha,\beta\atop\alpha\neq\beta}\int d\mu_{\alpha\beta}(k)\bigg[(p_{\alpha}\cdot p_{\beta})\bigg\{(\ell_{\alpha}\cdot \ell_{\beta})(S_{\alpha}\wedge \ell_{\alpha})^{\mu}p_{\beta}^{\nu}+(k\cdot p_{\beta})(S_{\alpha}\wedge \ell_{\alpha})^{\mu}\ell_{\alpha}^{\nu}\nonumber \\
& +(\ell_{\alpha}\wedge \ell_{\beta})_{\alpha}p_{\beta}^{\mu}l_{\alpha}^{\nu}+(\ell_{\alpha}\wedge p_{\beta})_{\alpha}\ell_{\alpha}^{\mu}\ell_{\alpha}^{\nu}-\frac{1}{2}\ell_{\alpha}^{2}(\ell_{\alpha}\wedge p_{\beta})_{\alpha}\eta^{\mu\nu}\bigg\}-(k\cdot p_{\beta})^{2}(S_{\alpha}\wedge \ell_{\alpha})^{\mu}p_{\alpha}^{\nu}\nonumber \\
& +(k\cdot p_{\alpha})(k\cdot p_{\beta})(S_{\alpha}\wedge \ell_{\alpha})^{\mu}p_{\beta}^{\nu}+(k\cdot p_{\alpha})(\ell_{\alpha}\wedge p_{\beta})_{\alpha}p_{\beta}^{\mu}\ell_{\alpha}^{\nu}-(k\cdot p_{\beta})(\ell_{\alpha}\wedge p_{\beta})_{\alpha}p_{\alpha}^{\mu}\ell_{\alpha}^{\nu}\nonumber \\
& -(k\cdot p_{\beta})(\ell_{\alpha}\wedge \ell_{\beta})_{\alpha}p_{\alpha}^{\mu}p_{\beta}^{\nu}+(k\cdot p_{\alpha})(\ell_{\alpha}\wedge \ell_{\beta})_{\alpha}p_{\beta}^{\mu}p_{\beta}^{\nu}-(\ell_{\alpha}\cdot \ell_{\beta})(\ell_{\alpha}\wedge p_{\beta})_{\alpha}p_{\alpha}^{\mu}p_{\beta}^{\nu}\nonumber \\
& +\frac{m_{\beta}^{2}}{d-2}(S_{\alpha}\wedge \ell_{\alpha})^{\mu}p_{\alpha}^{\nu}+(\mu\leftrightarrow\nu)\bigg].\label{eq:grav3}
\end{align} 

The total canonically normalized graviton amplitude is then summarized as
\begin{align}
\mathcal{A}_{g}(k)\bigr\vert&_{\mathcal{O}(\eta^{3},S^1)}  =-\frac{i\epsilon^*_{\mu\nu}}{8m_{Pl}^{3(d-2)/2}}\sum_{\alpha,\beta\atop\alpha\neq\beta}\int d\mu_{\alpha\beta}(k)\bigg[\frac{(p_{\alpha}\cdot p_{\beta})(k\cdot \ell_{\beta})\ell_{\alpha}^{2}}{2(k\cdot p_{\alpha})^{2}}\Big\{\Big((\ell_{\beta}\wedge p_{\beta})_{\alpha}-(\ell_{\beta}\wedge p_{\alpha})_{\beta}\Big)p_{\alpha}^{\mu}p_{\alpha}^{\nu}\nonumber \\
& +(p_{\alpha}\cdot p_{\beta})(S_{\alpha}\wedge k)^{\mu}p_{\alpha}^{\nu}\Big\}-\frac{(p_{\alpha}\cdot p_{\beta})\ell_{\alpha}^{2}}{k\cdot p_{\alpha}}\bigg\{\frac{1}{2}(p_\alpha\cdot p_\beta)(S_{\alpha}\wedge k)^{\mu}\ell_{\alpha}^{\nu}+\frac{1}{2}(k\cdot p_{\beta})\Big((S_{\alpha}\wedge k)^{\mu}+(S_{\alpha}\wedge \ell_{\alpha})^{\mu}\Big)p_{\alpha}^{\nu}\nonumber \\
&+\frac{1}{2}(k\cdot \ell_{\beta})(S_{\alpha}\wedge p_{\beta})^{\mu}p_{\alpha}^{\nu} +\frac{1}{2}(\ell_{\alpha}\wedge \ell_{\beta})_{\beta}p_{\alpha}^{\mu}p_{\alpha}^{\nu}+\frac{1}{2}(\ell_{\alpha}\wedge \ell_{\beta})_{\alpha}p_{\alpha}^{\mu}p_{\beta}^{\nu}+\bigg((\ell_{\beta}\wedge p_{\beta})_{\alpha}-(\ell_{\beta}\wedge p_{\alpha})_{\beta}-\frac{1}{2}(k\wedge p_{\beta})_{\alpha}\bigg)p_{\alpha}^{\mu}\ell_{\beta}^{\nu}\bigg\}\nonumber \\
& -\frac{(k\cdot p_{\beta})\ell_{\alpha}^{2}}{2(k\cdot p_{\alpha})}\Big\{(\ell_{\beta}\wedge p_{\beta})_{\alpha}-(\ell_{\beta}\wedge p_{\alpha})_{\beta}\Big\} p_{\alpha}^{\mu}p_{\alpha}^{\nu}+(p_{\alpha}\cdot p_{\beta})\bigg\{\frac{\ell_{\alpha}^{2}}{2}(S_{\alpha}\wedge \ell_{\beta})^{\mu}p_{\beta}^{\nu}-(\ell_{\alpha}\cdot \ell_{\beta})(S_{\alpha}\wedge \ell_{\alpha})^{\mu}p_{\beta}^{\nu}\nonumber \\
& +(k\cdot p_{\beta})(S_{\alpha}\wedge \ell_{\alpha})^{\mu}\ell_{\alpha}^{\nu}+\frac{1}{2}\ell_{\alpha}^{2}(S_{\alpha}\wedge p_{\beta})^{\mu}\ell_{\beta}^{\nu}+(\ell_{\alpha}\wedge \ell_{\beta})_{\alpha}p_{\beta}^{\mu}\ell_{\alpha}^{\nu}+(\ell_{\alpha}\wedge p_{\beta})_{\alpha}\ell_{\alpha}^{\mu}\ell_{\alpha}^{\nu}-\frac{\ell_{\alpha}^{2}}{2}(\ell_{\alpha}\wedge p_{\beta})_{\alpha}\eta^{\mu\nu}\bigg\}\nonumber \\
& +\frac{1}{2}\Big\{ \ell_{\alpha}^{2}(\ell_{\beta}\wedge p_{\beta})_{\alpha}-\ell_{\beta}^{2}(\ell_{\alpha}\wedge p_{\beta})_{\alpha}\Big\} p_{\alpha}^{\mu}p_{\beta}^{\nu}+\bigg\{(k\cdot p_{\beta})(S_{\alpha}\wedge \ell_{\alpha})^{\mu}-\frac{\ell_{\alpha}^{2}}{2}(S_{\alpha}\wedge p_{\beta})^{\mu}\nonumber \\
& +(\ell_{\alpha}\wedge p_{\beta})_{\alpha}l_{\alpha}^{\mu}+(\ell_{\alpha}\wedge \ell_{\beta})_{\alpha}p_{\beta}^{\mu}\Bigg\}\Big((k\cdot p_{\alpha})p_{\beta}^{\nu}-(k\cdot p_{\beta})p_{\alpha}^{\nu}\Big)+(\mu\leftrightarrow\nu)\bigg].\label{eq:gravtot}
\end{align}
The on-shell difference between this and the double copy prediction Eq. (\ref{eq:gravitondc}) is
\begin{align}
\hat{\mathcal{A}}_{g}(k)- & \mathcal{A}_{g}(k)\bigr\vert_{\mathcal{O}(\eta^{3},S^{1})}=\frac{i\epsilon_{\mu\nu}^{*}}{8m_{Pl}^{3(d-2)/2}}\sum_{{\alpha,\beta\atop \alpha\neq\beta}}\int d\mu_{\alpha\beta}(k)\bigg[\frac{p_{\alpha}\cdot p_{\beta}}{2(k\cdot p_{\alpha})}\bigg\{-\frac{1}{2}\ell_{\alpha}^{2}(p_{\alpha}\cdot p_{\beta})(S_{\alpha}\wedge k)^{\mu}k^{\nu}-(\ell_{\alpha}\cdot\ell_{\beta})(\ell_{\alpha}\wedge p_{\beta})_{\alpha}k^{\mu}p_{\alpha}^{\nu}\nonumber\\
& -\frac{1}{2}\ell_{\alpha}^{2}\Big\{(\ell_{\beta}\wedge p_{\beta})_{\alpha}-(\ell_{\beta}\wedge p_{\alpha})_{\beta}\Big\} k{}^{\mu}p_{\alpha}^{\nu}\bigg\}+\frac{1}{2}(p_{\alpha}\cdot p_{\beta})\bigg\{(k\cdot p_{\beta})(S_{\alpha}\wedge\ell_{\alpha})^{\mu}k^{\nu}+\frac{1}{2}\ell_{\alpha}^{2}(S_{\alpha}\wedge p_{\beta})^{\mu}k^{\nu}\nonumber\\
& +(\ell_{\alpha}\wedge\ell_{\beta})_{\alpha}k^{\nu}p_{\beta}^{\mu}-\frac{1}{2}(\ell_{\alpha}\wedge p_{\beta})_{\alpha}k{}^{\mu}k{}^{\nu}+2(\ell_{\alpha}\wedge p_{\beta})_{\alpha}\ell_{\alpha}^{\mu}k{}^{\nu}-\ell_{\alpha}^{2}(l_{\alpha}\wedge p_{\beta})_{\alpha}\eta^{\mu\nu}\bigg\}\nonumber\\
&-\frac{1}{2}(\ell_{\alpha}\wedge p_{\beta})_{\alpha}\Big((k\cdot p_{\beta})p_{\alpha}^{\nu}-(k\cdot p_{\alpha})p_{\beta}^{\nu}\Big)k^{\mu}+(\mu\leftrightarrow\nu)\bigg].
\end{align}
We then use the de-Donder gauge condition to write this as
\begin{align}
\hat{\mathcal{A}}_{g}(k)- & \mathcal{A}_{g}(k)\bigr\vert_{\mathcal{O}(\eta^{3},S^{1})}=\frac{i\epsilon_{\mu\nu}^{*}}{8m_{Pl}^{3(d-2)/2}}\sum_{{\alpha,\beta\atop \alpha\neq\beta}}\int d\mu_{\alpha\beta}(k)\frac{\eta^{\mu\nu}}{4}\bigg[(p_{\alpha}\cdot p_{\beta})\bigg\{\Big(-2\ell_{\alpha}^{2}+2k\cdot\ell_{\alpha}-(\ell_{\alpha}\cdot\ell_{\beta})\Big)(\ell_{\alpha}\wedge p_{\beta})_{\alpha}\nonumber\\
& -\frac{1}{2}\ell_{\alpha}^{2}\Big((\ell_{\beta}\wedge p_{\beta})_{\alpha}-(\ell_{\beta}\wedge p_{\alpha})_{\beta}-(k\wedge p_{\beta})_{\alpha}\Big)+(k\cdot p_{\beta})\Big((k\wedge\ell_{\alpha})_{\alpha}+(\ell_{\alpha}\wedge\ell_{\beta})_{\alpha}\Big)\bigg\}+(\mu\leftrightarrow\nu)\bigg],
\end{align}
which can be easily shown to vanish on-shell. Thus, we have verified that the graviton amplitude is as predicted up to gauge terms.

\subsection{Dilaton radiation}
\label{sec:dilrad}
\begin{figure}
	\centering
	\includegraphics[scale=0.56]{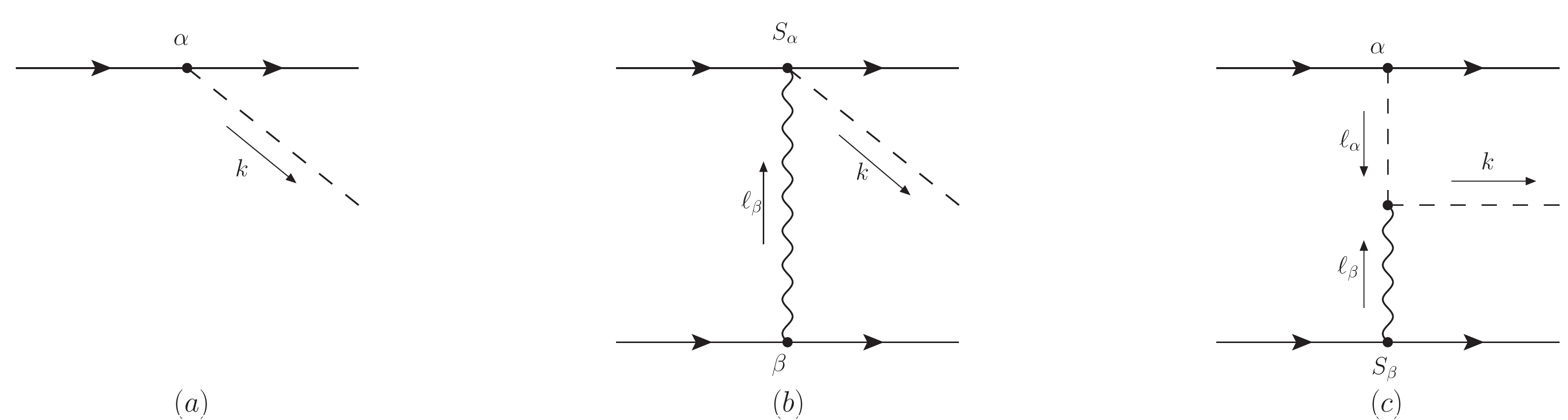}
		\caption{\label{fig:dilaton1pt} Feynman diagrams contributing to the dilaton source $\tilde{J}(k)$  at $\mathcal{O}(\eta^2)$, with a single spin insertion. As earlier, the dashed lines and the wavy lines represent the dilaton and the graviton respectively. Diagram $(a)$ represents corrections to the spin-independent piece of $\tilde{J}(k)$  due to the equations of motion. Diagrams $(b)-(c)$ correspond to a single spin insertion.}    
\end{figure}
In this subsection, we calculate the leading order dilaton radiation amplitude. Fig.~\ref{fig:dilaton1pt}(a) gives contributions to the dilaton amplitude from trajectory deflections,
\begin{align}
\tilde{J}(k)\bigr\vert_{\text{Fig.~\ref{fig:dilaton1pt}(a)};\mathcal{O}(\eta^2,S^1)}=&\sum_{\alpha}\int d\tau_\alpha e^{ik\cdot x_\alpha}\frac{i}{k\cdot v_\alpha}\bigg[\frac{k\cdot \dot{v}_\alpha}{k\cdot v_\alpha}v_\alpha\cdot p_{\alpha}-\dot{v}_\alpha\cdot p_\alpha-v_\alpha\cdot \dot{p}_\alpha\bigg]\Biggr\vert_{\mathcal{O}(\eta^{2},S^1)}.
\end{align}
Substituting the solutions to equations of motion in the above gives
\begin{align}
\tilde{J}(k)&\bigr\vert_{\text{Fig.~\ref{fig:dilaton1pt}(a)};\mathcal{O}(\eta^2,S^1)}  =\frac{i}{4m_{Pl}^{d-2}}\sum_{\alpha,\beta\atop \alpha\neq\beta}\int d\mu_{\alpha\beta}(k)\ell_{\alpha}^{2}\Bigg[\frac{p_{\alpha}\cdot p_{\beta}}{(k\cdot p_{\alpha})^{2}}p_{\alpha}^{2}(k\cdot \ell_{\beta})\Big\{(\ell_{\beta}\wedge p_{\alpha})_{\beta}\\
& -(\ell_{\beta}\wedge p_{\beta})_{\alpha}\Big\}+\frac{p_{\alpha}\cdot p_{\beta}}{k\cdot p_{\alpha}}p_{\alpha}^{2}(\ell_{\alpha}\wedge \ell_{\beta})_{\beta}+\frac{k\cdot p_{\beta}}{k\cdot p_{\alpha}}p_{\alpha}^{2}\Big\{(\ell_{\beta}\wedge p_{\beta})_{\alpha}-(\ell_{\beta}\wedge p_{\alpha})_{\beta}\Big\}+2(p_{\alpha}\cdot p_{\beta})(\ell_{\beta}\wedge p_{\alpha})_{\beta}\bigg].\label{eq:diltraj}
\end{align}

The other two contributions are calculated at zero deflections in the particle trajectories. Fig.~\ref{fig:dilaton1pt}(b) involving an intermediate graviton contributes
\begin{equation}
\tilde{J}(k)\bigr\vert_{\text{Fig.~\ref{fig:dilaton1pt}(b)};\mathcal{O}(\eta^2,S^1)}=-\frac{i}{2m_{Pl}^{d-2}}\sum_{\alpha,\beta\atop \alpha\neq\beta}\int d\mu_{\alpha\beta}(k)\ell_{\alpha}^{2}(p_{\alpha}\cdot p_{\beta})(\ell_{\beta}\wedge p_{\alpha})_{\beta}.
\end{equation}

Finally, we have Fig.~\ref{fig:dilaton1pt}(c) involving the 3-point graviton-dilaton vertex in the bulk, contributing
\begin{equation}
\tilde{J}(k)\bigr\vert_{\text{Fig.~\ref{fig:dilaton1pt}(c)};\mathcal{O}(\eta^2, S^1)}=\frac{i}{2m_{Pl}^{d-2}}\sum_{\alpha\beta\atop \alpha\neq\beta}\int d\mu_{\alpha,\beta}(k)p_{\alpha}^{2}(k\cdot p_{\beta})(\ell_{\alpha}\wedge \ell_{\beta})_{\beta}.\label{eq:dilbulk}
\end{equation}

The total dilaton amplitude is
\begin{align}
\mathcal{A}_{s}(k)\bigr\vert_{\mathcal{O}(\eta^3, S^1)}  =&-\frac{i}{8m_{Pl}^{3(d-2)/2}(d-2)^{1/2}}\sum_{\alpha,\beta\atop \alpha\neq\beta}\int d\mu_{\alpha\beta}(k)p_{\alpha}^{2}\bigg[\frac{p_{\alpha}\cdot p_{\beta}}{(k\cdot p_{\alpha})^{2}}\ell_{\alpha}^{2}(k\cdot \ell_{\beta})\Big\{(\ell_{\beta}\wedge p_{\alpha})_{\beta}-(\ell_{\beta}\wedge p_{\beta})_{\alpha}\Big\}\nonumber \\
 &+\frac{k\cdot p_{\beta}}{k\cdot p_{\alpha}}\ell_{\alpha}^{2}\Big\{(\ell_{\beta}\wedge p_{\beta})_{\alpha}-(\ell_{\beta}\wedge p_{\alpha})_{\beta}\Big\}+\frac{p_{\alpha}\cdot p_{\beta}}{k\cdot p_{\alpha}}\ell_{\alpha}^{2}(\ell_{\alpha}\wedge \ell_{\beta})_{\beta}+2(k\cdot p_{\beta})(\ell_{\alpha} \ell_{\beta})_{\beta}\bigg],
\end{align}
as predicted by the double copy Eq. (\ref{eq:dilatondc}). 

\section{Discussions and Conclusions}
\label{sec:Disc}
Lessons learnt from the study of scattering amplitudes in quantum field theory and string theory are proving useful in the study of classical gravity. Some examples of this are the use of on-shell methods \cite{Neill:2013wsa,Bjerrum-Bohr:2013bxa}, leading singularities \cite{LeadingSing} and soft theorems \cite{Laddha:2018rle}. In this paper, we continued exploring the idea of using the BCJ double copy to obtain perturbative classical gravitational radiation \cite{Goldberger:2016iau}.
Specifically, we have established a complete correspondence between perturbative classical Yang-Mills and gravitational radiation, emitted by spinning sources, up to linear order in spin. Consistency with the double copy fixes the couplings of the spinning sources to the Yang-Mills field. The double copied field is decomposed into its antisymmetric, symmetric-traceless and trace components. These are the predictions of the double copy for  the axion, graviton and dilaton radiation fields emitted by spinning sources in a gravitational theory. The double copy also fixes the gravitational action in the bulk to be the same as the low energy effective action of oriented closed strings Eq. (\ref{eq:sb}) and predicts a linear interaction of the spins with the axion Eq. (\ref{eq:SBaction}). We explicitly calculated the graviton, dilaton and axion radiation amplitudes in this theory (see \cite{JV} for analogous pure gravity results), to linear order in spin, and showed that they are exactly as predicted from the double copy. 

By including spin as a dynamical degree of freedom for our sources, we have brought the classical double copy closer to being useful for the computation of gravitational radiation from astrophysically relevant sources. Our calculation can be used to generate gravitational radiation from sources along arbitrary trajectories. Hence they can be used to compute gravitational scattering as well as radiation from objects in bound orbits. 
However, the gravitational radiation so obtained is in a theory which includes the dilaton and the axion in addition to the graviton. In order to arrive at the corresponding results in pure gravity, we need a systematic method that cancels out the effect of these extra degrees of freedom.
For spinless sources, Ref.~\cite{Luna:2017dtq} proposed a way of removing the contribution of the dilaton at leading order in perturbation, by  adding an appropriate ghost scalar to the Yang-Mills side, in a manner similar to \cite{Johansson:2014zca}.
% Another approach to obtaining pure gravity results could be to work in $d=4$, where one can obtain the pure gravity Einstein-Hilbert action by directly squaring the Yang-Mills one \cite{Ananth:2007zy}. 
Once we find a method of obtaining pure gravity results, it would be useful to write down an effective theory for the Yang-Mills sources, that upon double copying, directly reproduces the effective field theory in \cite{Goldberger:2004jt,Porto:2005ac} which treats the binary inspiral problem in a Post-Newtonian expansion.

For $d=4$, the leading order metric sourced by a single, static particle at the origin, with angular momentum $J$ about the $z$-axis, has a spin dependent piece given by $h_{t\phi}\bigr\vert_{\mathcal{O}(J^1)}=-\frac{2G_N J\sin ^2 \theta}{r}. $
We recall that the non-spinning part of the metric at this order in perturbation matches with the Schwarzschild solution \cite{Goldberger:2016iau}. At leading order in perturbation, there are no contributions from the dilaton or the axion to the metric. Hence, as expected, we find that the static, single particle limit of the metric matches that of the Kerr blackhole at this order. This indicates, that the gravitational solutions are spinning blackholes, albeit with dilatonic and axionic hair. It would be very interesting to see if there they are related to blackhole solutions in string theory.
%, for example, the ones in \cite{Campbell:1990ai}.

In order to be relevant for gravitational wave detectors, the double copy correspondence needs to be extended to higher orders, both in the coupling strength and in spin. We would also need to include finite size corrections to the sources by including higher order operators on the worldline that are allowed by the symmetries. The correspondence can be extended to charged, spinning sources. A further simplification of the results in this paper can be achieved by finding the analogous bi-adjoint scalar theory \cite{Cachazo:2013iea,BDC,Goldberger:2017frp}, the double copy of whose solution would produce the Yang-Mills radiation calculated here. A different approach to simplifying the perturbative expansion of the Einstein-Hilbert Lagrangian is by direct factorization of the action \cite{Bern:1999ji,Cheung:2016say}. It would be interesting to see if it is related to the classical double copy as described here. Another direction for future research is to explore the class of solutions of Einstein's equations that can be generated by this correspondence. Finally, the theoretical significance of the classical double copy for Einstein's general relativity remains an exciting and open question.

\section{Acknowledgements}
We gratefully acknowledge Walter Goldberger for suggesting this problem, for many valuable discussions and collaboration, and also for comments on this manuscript. SP also thanks Ghanashyam Date, Brian Henning, Dileep Jatkar, Alok Laddha, R. Loganayagam, Shiraz Minwalla, Ashoke Sen, Junpu Wang for useful discussions; and Brian Henning for a careful reading of an earlier version of this manuscript.


\begin{references}
	

	
	\bibitem{Bern:2008qj} 
	Z.~Bern, J.~J.~M.~Carrasco and H.~Johansson,
	%New Relations for Gauge-Theory Amplitudes,"%
	Phys.\ Rev.\ D {\bf 78}, 085011 (2008).
	%doi:10.1103/PhysRevD.78.085011
	%  [arXiv:0805.3993 [hep-ph]].
	%%CITATION = doi:10.1103/PhysRevD.78.085011;%%
	%409 citations counted in INSPIRE as of 17 Dec 2016
	
	
	\bibitem{Bern:2010ue} 
	Z.~Bern, J.~J.~M.~Carrasco and H.~Johansson,
	%``Perturbative Quantum Gravity as a Double Copy of Gauge Theory,''
	Phys.\ Rev.\ Lett.\  {\bf 105}, 061602 (2010).
	%  [arXiv:1004.0476 [hep-th]].
	%%CITATION = doi:10.1103/PhysRevLett.105.061602;%%
	%261 citations counted in INSPIRE as of 19 May 2017
	
	\bibitem{Bern:2010yg} 
	Z.~Bern, T.~Dennen, Y.~T.~Huang and M.~Kiermaier,
	%``Gravity as the Square of Gauge Theory,''
	Phys.\ Rev.\ D {\bf 82}, 065003 (2010).
	%doi:10.1103/PhysRevD.82.065003
	%[arXiv:1004.0693 [hep-th]].
	%%CITATION = doi:10.1103/PhysRevD.82.065003;%%
	%187 citations counted in INSPIRE as of 17 Dec 2016
	
	
	\bibitem{Kawai:1985xq} 
	H.~Kawai, D.~C.~Lewellen and S.~H.~H.~Tye,
	%``A Relation Between Tree Amplitudes of Closed and Open Strings,''
	Nucl.\ Phys.\ B {\bf 269}, 1 (1986).
	%doi:10.1016/0550-3213(86)90362-7
	%%CITATION = doi:10.1016/0550-3213(86)90362-7;%%
	%548 citations counted in INSPIRE as of 17 Dec 2016
	
	\bibitem{Loops}
	Z. Bern, J. J. M. Carrasco, L. J. Dixon, H. Johansson, and R.
	Roiban, %Simplifying multiloop integrands and ultraviolet divergences of gauge theory and gravity amplitudes, 
	Phys. Rev. D {\bf 85}, 105014 (2012); Z. Bern, C. Boucher-Veronneau, and H. Johansson, $\mathcal{N}\geq 4$
	%supergravity amplitudes from gauge theory at one loop,
	Phys. Rev. D {\bf84}, 105035 (2011); C. Boucher-Veronneau
	and L. J. Dixon, %$\mathcal{N}\geq 4$ supergravity amplitudes from gauge theory at two loops, 
	J. High Energy Phys. 12 046 (2011)
	; J. J. M. Carrasco and H. Johansson, %Five-point amplitudes in $\mathcal{N}= 4$ super-Yang-Mills theory and $\mathcal{N}=8$ supergravity,
	Phys. Rev. D {\bf 85}, 025006 (2012); Z. Bern, S. Davies, T.
	Dennen, Y. T. Huang, and J. Nohle, %Color-kinematics duality for pure Yang-Mills and gravity at one and two loops, 
	Phys. Rev. D {\bf 92}, 045041 (2015); J. J. M. Carrasco, M. Chiodaroli, M. Gunaydin, and R.
	Roiban, %One-loop four-point amplitudes in pure and matter coupled N <¼ 4 supergravity, 
	J. High Energy Phys. 03 056
	(2013);
	M. Chiodaroli, M. Gunaydin, H. Johansson,
	and R. Roiban,
	%Scattering amplitudes in N ¼ 2 MaxwellEinstein and Yang-Mills/Einstein supergravity,
	J. High
	Energy Phys. 01 081 (2015);
	M. Chiodaroli, M.
	Gunaydin, H. Johansson, and R. Roiban, %Complete Construction of Magical, Symmetric and Homogeneous N ¼ 2 Supergravities as Double Copies of Gauge Theories,
	Phys. Rev. Lett. {\bf117}, 011603 (2016);
	Z.~Bern, J.~J.~M.~Carrasco, W.~M.~Chen, H.~Johansson, R.~Roiban and M.~Zeng,
	%``The Five-Loop Four-Point Integrand of N=8 Supergravity as a Generalized Double Copy,''
	arXiv:1708.06807 [hep-th];
	%%CITATION = ARXIV:1708.06807;%%
	%6 citations counted in INSPIRE as of 21 Dec 2017
	Z.~Bern, J.~J.~Carrasco, W.~M.~Chen, H.~Johansson and R.~Roiban,
	%``Gravity Amplitudes as Generalized Double Copies of Gauge-Theory Amplitudes,''
	Phys.\ Rev.\ Lett.\  {\bf 118}, no. 18, 181602 (2017).
	% [arXiv:1701.02519 [hep-th]].
	%%CITATION = doi:10.1103/PhysRevLett.118.181602;%%
	%16 citations counted in INSPIRE as of 21 Dec 2017

	
	
	
	\bibitem{Carrasco:2015iwa} 
	J.~J.~M.~Carrasco,
	%``Gauge and Gravity Amplitude Relations,''
	% doi:10.1142/9789814678766_0011
	arXiv:1506.00974 [hep-th].
	%%CITATION = doi:10.1142/9789814678766_0011;%%
	%20 citations counted in INSPIRE as of 17 Dec 2016

	

	
	\bibitem{Monteiro:2014cda} 
	R.~Monteiro, D.~O'Connell and C.~D.~White,
	%``Black holes and the double copy,''
	JHEP {\bf 1412}, 056 (2014).
	%doi:10.1007/JHEP12(2014)056
	% [arXiv:1410.0239 [hep-th]].
	%%CITATION = doi:10.1007/JHEP12(2014)056;%%
	%35 citations counted in INSPIRE as of 17 Dec 2016
	
	
	\bibitem{KS2}
	A.~Luna, R.~Monteiro, D.~O'Connell and C.~D.~White,
	%``The classical double copy for TaubÐNUT spacetime,''
	Phys.\ Lett.\ B {\bf 750}, 272 (2015).
	%%CITATION = doi:10.1016/j.physletb.2015.09.021;%%
	%39 citations counted in INSPIRE as of 25 Nov 2017
	
	
	\bibitem{KS3} 
	A.~Luna, R.~Monteiro, I.~Nicholson, D.~O'Connell and C.~D.~White,
	%``The double copy: Bremsstrahlung and accelerating black holes,''
	JHEP {\bf 1606}, 023 (2016).
	%%CITATION = doi:10.1007/JHEP06(2016)023;%%
	%28 citations counted in INSPIRE as of 25 Nov 2017
	
	
	
	\bibitem{KSothers}
	A.~K.~Ridgway and M.~B.~Wise,
	%``Static Spherically Symmetric Kerr-Schild Metrics and Implications for the Classical Double Copy,''
	Phys.\ Rev.\ D {\bf 94}, no. 4, 044023 (2016);
	%%CITATION = doi:10.1103/PhysRevD.94.044023;%%
	G.~L.~Cardoso, S.~Nagy and S.~Nampuri,
	%``A double copy for $ \mathcal{N}=2 $ supergravity: a linearised tale told on-shell,''
	JHEP {\bf 1610}, 127 (2016);
	%  [arXiv:1609.05022 [hep-th]];
	%%CITATION = doi:10.1007/JHEP10(2016)127;%%
	%13 citations counted in INSPIRE as of 21 Dec 2017
	G.~Cardoso, S.~Nagy and S.~Nampuri,
	%``Multi-centered $ \mathcal{N}=2 $ BPS black holes: a double copy description,''
	JHEP {\bf 1704}, 037 (2017);
	%[arXiv:1611.04409 [hep-th]].
	%%CITATION = doi:10.1007/JHEP04(2017)037;%%
	%15 citations counted in INSPIRE as of 21 Dec 2017 
	Y.~Z.~Chu,
	%``More On Cosmological Gravitational Waves And Their Memories,''
	arXiv:1611.00018 [gr-qc];
	T.~Adamo, E.~Casali, L.~Mason and S.~Nekovar,
	%``Scattering on plane waves and the double copy,''
	Class.\ Quant.\ Grav.\  {\bf 35}, no. 1, 015004 (2018);
	%[arXiv:1706.08925 [hep-th]];
	%%CITATION = doi:10.1088/1361-6382/aa9961;%%
	%13 citations counted in INSPIRE as of 21 Dec 2017
	%%CITATION = ARXIV:1611.00018;%%
	N.~Bahjat-Abbas, A.~Luna and C.~D.~White,
	%``The Kerr-Schild double copy in curved spacetime,''
	arXiv:1710.01953 [hep-th];
	%%CITATION = ARXIV:1710.01953;%%
	M.~Carrillo-Gonzalez, R.~Penco and M.~Trodden,
	%``The classical double copy in maximally symmetric spacetimes,''
	arXiv:1711.01296 [hep-th].  
	
	
	\bibitem{Goldberger:2004jt} 
	W.~D.~Goldberger and I.~Z.~Rothstein,
	%``An Effective field theory of gravity for extended objects,''
	Phys.\ Rev.\ D {\bf 73}, 104029 (2006).
	%%CITATION = doi:10.1103/PhysRevD.73.104029;%%
	%220 citations counted in INSPIRE as of 22 Dec 2017
	
	
	\bibitem{Porto:2005ac} 
	R.~A.~Porto,
	%``Post-Newtonian corrections to the motion of spinning bodies in NRGR,''
	Phys.\ Rev.\ D {\bf 73}, 104031 (2006).
	%  [gr-qc/0511061].
	%%CITATION = doi:10.1103/PhysRevD.73.104031;%%
	%120 citations counted in INSPIRE as of 02 Aug 2017
	
	\bibitem{Porto:2016pyg} 
	R.~A.~Porto,
	%``The effective field theorist’s approach to gravitational dynamics,''
	Phys.\ Rept.\  {\bf 633}, 1 (2016)
	doi:10.1016/j.physrep.2016.04.003
	[arXiv:1601.04914 [hep-th]].
	%%CITATION = doi:10.1016/j.physrep.2016.04.003;%%
	%31 citations counted in INSPIRE as of 16 Feb 2018
	
	
	\bibitem{Goldberger:2016iau} 
	W.~D.~Goldberger and A.~K.~Ridgway,
	%``Radiation and the classical double copy for color charges,''
	Phys.\ Rev.\ D {\bf 95}, no. 12, 125010 (2017).
	%[arXiv:1611.03493 [hep-th]].
	%%CITATION = doi:10.1103/PhysRevD.95.125010;%%
	%27 citations counted in INSPIRE as of 21 Dec 2017   
	
	
	\bibitem{Goldberger:2017frp} 
	W.~D.~Goldberger, S.~G.~Prabhu and J.~O.~Thompson,
	%``Classical gluon and graviton radiation from the bi-adjoint scalar double copy,''
	Phys.\ Rev.\ D {\bf 96}, no. 6, 065009 (2017).
	%  doi:10.1103/PhysRevD.96.065009
	%%CITATION = doi:10.1103/PhysRevD.96.065009;%%
	%13 citations counted in INSPIRE as of 23 Dec 2017
	
	
	\bibitem{Goldberger:2017vcg} 
	W.~D.~Goldberger and A.~K.~Ridgway,
	%``Bound states and the classical double copy,''
	arXiv:1711.09493 [hep-th].
	%%CITATION = ARXIV:1711.09493;%%
	

	
	
	\bibitem{TheLIGOScientific:2017qsa} 
	B.~P.~Abbott {\it et al.} [LIGO Scientific and Virgo Collaborations],
	%``GW170817: Observation of Gravitational Waves from a Binary Neutron Star Inspiral,''
	Phys.\ Rev.\ Lett.\  {\bf 119}, no. 16, 161101 (2017).
	%[arXiv:1710.05832 [gr-qc]].
	%%CITATION = doi:10.1103/PhysRevLett.119.161101;%%
	%269 citations counted in INSPIRE as of 21 Dec 2017
	
	
	
	\bibitem{Luna:2016hge} 
	A.~Luna, R.~Monteiro, I.~Nicholson, A.~Ochirov, D.~O'Connell, N.~Westerberg and C.~D.~White,
	%``Perturbative spacetimes from Yang-Mills theory,''
	JHEP {\bf 1704}, 069 (2017).
	% [arXiv:1611.07508 [hep-th]].
	%%CITATION = doi:10.1007/JHEP04(2017)069;%%
	%23 citations counted in INSPIRE as of 21 Dec 2017
	
		\bibitem{Goldberger:2017axi}
	W.~D.~Goldberger, J.~Li and S.~G.~Prabhu, arXiv:1712.09250 [hep-th].
	
	
	
	\bibitem{Chester:2017} D. Chester, arXiv:1712.08684 [hep-th].
	
	
	
	
	\bibitem{Mathisson:1937zz} 
	M.~Mathisson,
	%``Neue mechanik materieller systemes,''
	Acta Phys.\ Polon.\  {\bf 6}, 163 (1937),
	%%CITATION = APPOA,6,163;%%
	%309 citations counted in INSPIRE as of 15 Feb 2018
	English translation: Gen Relativ Gravit (2010) 42: 1011. https://doi.org/10.1007/s10714-010-0939-y
	
	\bibitem{Papapetrou:1951pa} 
	A.~Papapetrou,
	%``Spinning test particles in general relativity. 1.,''
	Proc.\ Roy.\ Soc.\ Lond.\ A {\bf 209}, 248 (1951).
	%%CITATION = doi:10.1098/rspa.1951.0200;%%
	%522 citations counted in INSPIRE as of 02 Aug 2017
	
	\bibitem{Dixon:1970zza} 
	W.~G.~Dixon,
	%``Dynamics of extended bodies in general relativity. I. Momentum and angular momentum,''
	Proc.\ Roy.\ Soc.\ Lond.\ A {\bf 314}, 499 (1970).
	%%CITATION = doi:10.1098/rspa.1970.0020;%%
	%219 citations counted in INSPIRE as of 02 Aug 2017
	
	\bibitem{Hanson:1974qy} 
	A.~J.~Hanson and T.~Regge,
	%``The Relativistic Spherical Top,''
	Annals Phys.\  {\bf 87}, 498 (1974).
	%%CITATION = doi:10.1016/0003-4916(74)90046-3;%%
	%210 citations counted in INSPIRE as of 02 Aug 2017
	
	
	
	\bibitem{Bailey:1975fe} 
	I.~Bailey and W.~Israel,
	%``Lagrangian Dynamics of Spinning Particles and Polarized Media in General Relativity,''
	Commun.\ Math.\ Phys.\  {\bf 42}, 65 (1975).
	%%CITATION = doi:10.1007/BF01609434;%%
	%72 citations counted in INSPIRE as of 02 Aug 2017
	
	%	\bibitem{Mino:1995fm} 
	%	Y.~Mino, M.~Shibata and T.~Tanaka,
	%	%``Gravitational waves induced by a spinning particle falling into a rotating black hole,''
	%	Phys.\ Rev.\ D {\bf 53}, 622 (1996)
	%	Erratum: [Phys.\ Rev.\ D {\bf 59}, 047502 (1999)];
	%	%%CITATION = doi:10.1103/PhysRevD.53.622, 10.1103/PhysRevD.59.047502;%%
	%	%38 citations counted in INSPIRE as of 02 Aug 2017
	%	T.~Tanaka, Y.~Mino, M.~Sasaki and M.~Shibata,
	%	%``Gravitational waves from a spinning particle in circular orbits around a rotating black hole,''
	%	Phys.\ Rev.\ D {\bf 54}, 3762 (1996).
	%	%%CITATION = doi:10.1103/PhysRevD.54.3762;%%
	%	%59 citations counted in INSPIRE as of 02 Aug 2017
	
	
	
	%	\bibitem{Delacretaz:2014oxa} 
	%	L.~V.~Delacretaz, S.~Endlich, A.~Monin, R.~Penco and F.~Riva,
	%	%``(Re-)Inventing the Relativistic Wheel: Gravity, Cosets, and Spinning Objects,''
	%	JHEP {\bf 1411}, 008 (2014)
	%	%[arXiv:1405.7384 [hep-th]].
	%	%%CITATION = doi:10.1007/JHEP11(2014)008;%%
	%	%25 citations counted in INSPIRE as of 18 Dec 2017
	
	%  
	%  \bibitem{Wald:1972sz} 
	%  R.~M.~Wald,
	%  %``Gravitational spin interaction,''
	%  Phys.\ Rev.\ D {\bf 6}, 406 (1972).
	%  doi:10.1103/PhysRevD.6.406
	%  %%CITATION = doi:10.1103/PhysRevD.6.406;%%
	%  %142 citations counted in INSPIRE as of 02 Aug 2017
	%  
	
	%\cite{Cachazo:2017jef}
	
	
	
	
	\bibitem{Porto+Rothstein} 
	R.~A.~Porto and I.~Z.~Rothstein,
	%``The Hyperfine Einstein-Infeld-Hoffmann potential,''
	Phys.\ Rev.\ Lett.\  {\bf 97}, 021101 (2006);
	%  [gr-qc/0604099].
	%%CITATION = doi:10.1103/PhysRevLett.97.021101;%%
	%102 citations counted in INSPIRE as of 02 Aug 2017
	%R.~A.~Porto and I.~Z.~Rothstein,
	%``Spin(1)Spin(2) Effects in the Motion of Inspiralling Compact Binaries at Third Order in the Post-Newtonian Expansion,''
	Phys.\ Rev.\ D {\bf 78}, 044012 (2008)
	Erratum: [Phys.\ Rev.\ D {\bf 81}, 029904 (2010)];
	%[arXiv:0802.0720 [gr-qc]].
	%%CITATION = doi:10.1103/PhysRevD.78.044012, 10.1103/PhysRevD.81.029904;%%
	%109 citations counted in INSPIRE as of 02 Aug 2017
	%R.~A.~Porto and I.~Z.~Rothstein,
	%``Next to Leading Order Spin(1)Spin(1) Effects in the Motion of Inspiralling Compact Binaries,''
	Phys.\ Rev.\ D {\bf 78}, 044013 (2008)
	Erratum: [Phys.\ Rev.\ D {\bf 81}, 029905 (2010)].
	%  doi:10.1103/PhysRevD.81.029905, 10.1103/PhysRevD.78.044013
	%[arXiv:0804.0260 [gr-qc]].
	%%CITATION = doi:10.1103/PhysRevD.81.029905, 10.1103/PhysRevD.78.044013;%%
	%114 citations counted in INSPIRE as of 23 Dec 2017
	
	
	\bibitem{sikivie}
	P.~Sikivie and N.~Weiss,
	%``Classical {Yang-Mills} Theory in the Presence of External Sources,''
	Phys.\ Rev.\ D {\bf 18}, 3809 (1978).
	%%CITATION = doi:10.1103/PhysRevD.18.3809;%%
	%106 citations counted in INSPIRE as of 04 Oct 2016
	
	
	\bibitem{Kalb:1974yc} 
	M.~Kalb and P.~Ramond,
	%``Classical direct interstring action,''
	Phys.\ Rev.\ D {\bf 9}, 2273 (1974).
	%%CITATION = doi:10.1103/PhysRevD.9.2273;%%
	%733 citations counted in INSPIRE as of 21 Dec 2017
	
	\bibitem{Pryce:1948sp}
	M. H. L. Pryce,
	%``The mass-centre in the restricted theory of relativity and its connexion with the quantum theory of elementary particles,''
	Proc.\ Roy.\ Soc.\ Lond.\ A {\bf 195}, 1040 (1948).
	%Proc. R. Soc. Lond. A 1948 195 62-81; DOI: 10.1098/rspa.1948.0103. Published 12 November 1948
	
	
	\bibitem{Ferrara:1992yc} 
	S.~Ferrara, M.~Porrati and V.~L.~Telegdi,
	%``g = 2 as the natural value of the tree level gyromagnetic ratio of elementary particles,''
	Phys.\ Rev.\ D {\bf 46}, 3529 (1992).
	%%CITATION = doi:10.1103/PhysRevD.46.3529;%%
	%144 citations counted in INSPIRE as of 18 Dec 2017
	
	\bibitem{Holstein:2006ry} 
	B.~R.~Holstein,
	%``Factorization in graviton scattering and the 'natural' value of the g factor,''
	Phys.\ Rev.\ D {\bf 74}, 085002 (2006).
	%%CITATION = doi:10.1103/PhysRevD.74.085002;%%
	%13 citations counted in INSPIRE as of 19 Dec 2017
	
	
	\bibitem{Scherk:1974mc} 
	J.~Scherk and J.~H.~Schwarz,
	%``Dual Models and the Geometry of Space-Time,''
	Phys.\ Lett.\  {\bf 52B}, 347 (1974).
	%%CITATION = doi:10.1016/0370-2693(74)90059-8;%%
	%205 citations counted in INSPIRE as of 18 Dec 2017
	
	\bibitem{Gross:1987hs}
	D.~J.~Gross, J.~H.~Sloan,
	%``The quartic effective action for the heterotic string,''
	Nucl. Phys. B {\bf291}, 41-89 (1987).
	
	
	\bibitem{Bern:2015uc}
	Z.~Bern, S.~Davies, J.~Nohle,
	Phys. Rev. D {\bf 93}, 105015 (2016).
	
	
	
	\bibitem{BgGauge} 
	B.~S.~DeWitt,
	%``Quantum Theory of Gravity. 2. The Manifestly Covariant Theory,''
	Phys.\ Rev.\  {\bf 162}, 1195 (1967);
	%doi:10.1103/PhysRev.162.1195
	%%CITATION = doi:10.1103/PhysRev.162.1195;%%
	%1313 citations counted in INSPIRE as of 17 Feb 2018
	L.~F.~Abbott,
	%``The Background Field Method Beyond One Loop,''
	Nucl.\ Phys.\ B {\bf 185}, 189 (1981).
	%doi:10.1016/0550-3213(81)90371-0
	%%CITATION = doi:10.1016/0550-3213(81)90371-0;%%
	%1082 citations counted in INSPIRE as of 17 Feb 2018
	

%\cite{Donoghue:1994dn}
\bibitem{Donoghue:1994dn} 
J.~F.~Donoghue,
%``General relativity as an effective field theory: The leading quantum corrections,''
Phys.\ Rev.\ D {\bf 50}, 3874 (1994)
doi:10.1103/PhysRevD.50.3874
[gr-qc/9405057].
%%CITATION = doi:10.1103/PhysRevD.50.3874;%%
%616 citations counted in INSPIRE as of 06 Mar 2018	
	
	\bibitem{Bjerrum-Bohr:2013bxa} 
	N.~E.~J.~Bjerrum-Bohr, J.~F.~Donoghue and P.~Vanhove,
	%``On-shell Techniques and Universal Results in Quantum Gravity,''
	JHEP {\bf 1402}, 111 (2014)
	doi:10.1007/JHEP02(2014)111
	[arXiv:1309.0804 [hep-th]].
	%%CITATION = doi:10.1007/JHEP02(2014)111;%%
	%31 citations counted in INSPIRE as of 16 Feb 2018
	
	\bibitem{Neill:2013wsa} 
	D.~Neill and I.~Z.~Rothstein,
	%``Classical Space-Times from the S Matrix,''
	Nucl.\ Phys.\ B {\bf 877}, 177 (2013)
	%doi:10.1016/j.nuclphysb.2013.09.007
	[arXiv:1304.7263 [hep-th]].
	%%CITATION = doi:10.1016/j.nuclphysb.2013.09.007;%%
	%19 citations counted in INSPIRE as of 16 Feb 2018
	
	\bibitem{LeadingSing} 
	F.~Cachazo and A.~Guevara,
	%``Leading Singularities and Classical Gravitational Scattering,''
	arXiv:1705.10262 [hep-th];
	%%CITATION = ARXIV:1705.10262;%%
	%3 citations counted in INSPIRE as of 16 Feb 2018
	A.~Guevara,
	%``Holomorphic Classical Limit for Spin Effects in Gravitational and Electromagnetic Scattering,''
	arXiv:1706.02314 [hep-th].
	%%CITATION = ARXIV:1706.02314;%%
	%4 citations counted in INSPIRE as of 16 Feb 2018
	
	
	\bibitem{Laddha:2018rle} 
	A.~Laddha and A.~Sen,
	%``Gravity Waves from Soft Theorem in General Dimensions,''
	arXiv:1801.07719 [hep-th].
	%%CITATION = ARXIV:1801.07719;%%
	
		\bibitem{JV}
	D.~Bini, A.~Geralico and J.~Vines,
	%``Hyperbolic scattering of spinning particles by a Kerr black hole,''
	Phys.\ Rev.\ D {\bf 96}, no. 8, 084044 (2017);
	% [arXiv:1707.09814 [gr-qc]];
	%%CITATION = doi:10.1103/PhysRevD.96.084044;%%
	%1 citations counted in INSPIRE as of 21 Dec 2017
	J.~Vines,
	%``Scattering of two spinning black holes in post-Minkowskian gravity, to all orders in spin, and effective-one-body mappings,''
	arXiv:1709.06016 [gr-qc].
	%%CITATION = ARXIV:1709.06016;%%
	%2 citations counted in INSPIRE as of 21 Dec 2017
	
	\bibitem{Luna:2017dtq} 
	A.~Luna, I.~Nicholson, D.~O'Connell and C.~D.~White,
	%``Inelastic Black Hole Scattering from Charged Scalar Amplitudes,''
	arXiv:1711.03901 [hep-th].
	%%CITATION = ARXIV:1711.03901;%%
	%3 citations counted in INSPIRE as of 21 Dec 2017
	
	\bibitem{Johansson:2014zca} 
	H.~Johansson and A.~Ochirov,
	%``Pure Gravities via Color-Kinematics Duality for Fundamental Matter,''
	JHEP {\bf 1511}, 046 (2015).
	% [arXiv:1407.4772 [hep-th]].
	%%CITATION = doi:10.1007/JHEP11(2015)046;%%
	
	
%	\bibitem{Ananth:2007zy} 
%	S.~Ananth and S.~Theisen,
%	%``KLT relations from the Einstein-Hilbert Lagrangian,''
%	Phys.\ Lett.\ B {\bf 652}, 128 (2007)
%	%doi:10.1016/j.physletb.2007.07.003
%	[arXiv:0706.1778 [hep-th]].
%	%%CITATION = doi:10.1016/j.physletb.2007.07.003;%%
%	%53 citations counted in INSPIRE as of 16 Feb 2018
	
%	
%	\bibitem{Campbell:1990ai} 
%	B.~A.~Campbell, M.~J.~Duncan, N.~Kaloper and K.~A.~Olive,
%	%``Axion hair for Kerr black holes,''
%	Phys.\ Lett.\ B {\bf 251}, 34 (1990);
%	%%CITATION = doi:10.1016/0370-2693(90)90227-W;%%
%	%61 citations counted in INSPIRE as of 23 Dec 2017
%	B.~A.~Campbell, N.~Kaloper and K.~A.~Olive,
%	%``Classical hair for Kerr-Newman black holes in string gravity,''
%	Phys.\ Lett.\ B {\bf 285}, 199 (1992);
%	%%CITATION = doi:10.1016/0370-2693(92)91452-F;%%
%	%44 citations counted in INSPIRE as of 23 Dec 2017
%	B.~A.~Campbell, N.~Kaloper, R.~Madden and K.~A.~Olive,
%	%``Physical properties of four-dimensional superstring gravity black hole solutions,''
%	Nucl.\ Phys.\ B {\bf 399}, 137 (1993).
%	%%CITATION = doi:10.1016/0550-3213(93)90620-5;%%
%	%28 citations counted in INSPIRE as of 23 Dec 2017
%	
	
	
	
	
	
	
	
	
	
	
	
	
	
	
	
	
	

	
	\bibitem{BDC} 
	C.~D.~White,
	%``Exact solutions for the biadjoint scalar field,''
	Phys.\ Lett.\ B {\bf 763}, 365 (2016);
	%doi:10.1016/j.physletb.2016.10.052
	%%CITATION = doi:10.1016/j.physletb.2016.10.052;%%
	%7 citations counted in INSPIRE as of 17 Dec 2016
	P.~J.~De Smet and C.~D.~White,
	%``Extended solutions for the biadjoint scalar field,''
	Phys.\ Lett.\ B {\bf 775}, 163 (2017).
	%  doi:10.1016/j.physletb.2017.11.007
	%[arXiv:1708.01103 [hep-th]].
	%%CITATION = doi:10.1016/j.physletb.2017.11.007;%%
	%2 citations counted in INSPIRE as of 23 Dec 2017

	
	
	
	\bibitem{Cachazo:2013iea} 
	F.~Cachazo, S.~He and E.~Y.~Yuan,
	%``Scattering of Massless Particles: Scalars, Gluons and Gravitons,''
	JHEP {\bf 1407}, 033 (2014)
	%doi:10.1007/JHEP07(2014)033
	[arXiv:1309.0885 [hep-th]].
	%%CITATION = doi:10.1007/JHEP07(2014)033;%%
	
	
	\bibitem{Bern:1999ji} 
	Z.~Bern and A.~K.~Grant,
	%``Perturbative gravity from QCD amplitudes,''
	Phys.\ Lett.\ B {\bf 457}, 23 (1999).
	% [hep-th/9904026].
	%%CITATION = doi:10.1016/S0370-2693(99)00524-9;%%
	%77 citations counted in INSPIRE as of 18 Dec 2017
	
	
	
	%  \bibitem{Bern:2015ooa} 
	%  Z.~Bern, S.~Davies and J.~Nohle,
	%  %``Double-Copy Constructions and Unitarity Cuts,''
	%  Phys.\ Rev.\ D {\bf 93}, no. 10, 105015 (2016)
	%  [arXiv:1510.03448 [hep-th]].
	%  %%CITATION = doi:10.1103/PhysRevD.93.105015;%%
	%  %7 citations counted in INSPIRE as of 27 Jan 2017
	%
	%
	%   
	% 
	%
	%  \bibitem{Cachazo:2013iea} 
	%  F.~Cachazo, S.~He and E.~Y.~Yuan,
	%  %``Scattering of Massless Particles: Scalars, Gluons and Gravitons,''
	%  JHEP {\bf 1407}, 033 (2014)
	%  %doi:10.1007/JHEP07(2014)033
	%  [arXiv:1309.0885 [hep-th]].
	%  %%CITATION = doi:10.1007/JHEP07(2014)033;%%
	% 
	%
	%\bibitem{otherphi31}
	% Z.~Bern, A.~De Freitas and H.~L.~Wong,
	%  %``On the coupling of gravitons to matter,''
	%  Phys.\ Rev.\ Lett.\  {\bf 84}, 3531 (2000)
	%  [hep-th/9912033].
	%   %%CITATION = doi:10.1103/PhysRevLett.84.3531;%%
	%\bibitem{otherphi32}
	%  R.~Monteiro and D.~O'Connell,
	%  %``The Kinematic Algebra From the Self-Dual Sector,''
	%  JHEP {\bf 1107}, 007 (2011)
	%  [arXiv:1105.2565 [hep-th]].
	%   %%CITATION = doi:10.1007/JHEP07(2011)007;%%
	%\bibitem{otherphi33}
	%  Y.~J.~Du, B.~Feng and C.~H.~Fu,
	%  %``BCJ Relation of Color Scalar Theory and KLT Relation of Gauge Theory,''
	%  JHEP {\bf 1108}, 129 (2011)
	%  %doi:10.1007/JHEP08(2011)129
	%  [arXiv:1105.3503 [hep-th]].
	%   %%CITATION = doi:10.1007/JHEP08(2011)129;%%
	%  \bibitem{otherphi34}
	%N.~E.~J.~Bjerrum-Bohr, P.~H.~Damgaard, R.~Monteiro and D.~O'Connell,
	%  %``Algebras for Amplitudes,''
	%  JHEP {\bf 1206}, 061 (2012)
	%  %doi:10.1007/JHEP06(2012)061
	%  [arXiv:1203.0944 [hep-th]].  
	%   %%CITATION = doi:10.1007/JHEP06(2012)061;%%
	%    
	%  \bibitem{Anastasiou:2014qba} 
	%  A.~Anastasiou, L.~Borsten, M.~J.~Duff, L.~J.~Hughes and S.~Nagy,
	%  %``Yang-Mills origin of gravitational symmetries,''
	%  Phys.\ Rev.\ Lett.\  {\bf 113}, no. 23, 231606 (2014)
	%  %doi:10.1103/PhysRevLett.113.231606
	%  [arXiv:1408.4434 [hep-th]].
	%  %%CITATION = doi:10.1103/PhysRevLett.113.231606;%%
	%  %16 citations counted in INSPIRE as of 17 Dec 2016
	%  
	%   \bibitem{Cheung:2016prv} 
	%  C.~Cheung and C.~H.~Shen,
	%  %``Symmetry and Action for Flavor-Kinematics Duality,''
	%  Phys.\ Rev.\ Lett.\  {\bf 118}, no. 12, 121601 (2017)
	%  [arXiv:1612.00868 [hep-th]].
	%  %%CITATION = doi:10.1103/PhysRevLett.118.121601;%%
	%  %7 citations counted in INSPIRE as of 14 Apr 2017
	%  
	%  
	% 
	%
	%
	%
	%
	%
	%\bibitem{White:2016jzc} 
	%  C.~D.~White,
	%  %``Exact solutions for the biadjoint scalar field,''
	%  Phys.\ Lett.\ B {\bf 763}, 365 (2016)
	%  %doi:10.1016/j.physletb.2016.10.052
	%  [arXiv:1606.04724 [hep-th]].
	%  %%CITATION = doi:10.1016/j.physletb.2016.10.052;%%
	%  %7 citations counted in INSPIRE as of 17 Dec 2016
	%
	%
	%  
	%
	%
	%  
	%  
	%  
	%
	%
	%
	%   
	% 
	%
	%
	%
	%
	\bibitem{Cheung:2016say} 
	  C.~Cheung and G.~N.~Remmen,
	  %``Twofold Symmetries of the Pure Gravity Action,''
	  arXiv:1612.03927 [hep-th].
	  %%CITATION = ARXIv:1612.03927;%%  
	    C.~Cheung and G.~N.~Remmen,
	  %``Hidden Simplicity of the Gravity Action,''
	  arXiv:1705.00626 [hep-th].
	  %%CITATION = ARXIv:1705.00626;%%
	


	
	
	
	
	

	
	
	
	
	
	
	
	
\end{references}
\end{document}